\documentclass[aps,pra,twocolumn,amsmath,amssymb,showpacs,floatfix,superscriptaddress,footinbib]{revtex4-2}

\usepackage{times}
\usepackage{mdframed}

\usepackage{xcolor}
\usepackage[utf8x]{inputenc}
\usepackage[english]{babel}
\usepackage[T1]{fontenc}
\usepackage{lmodern}
\usepackage{amssymb}
\usepackage{amsthm}
\usepackage{amsmath}
\usepackage[pdftex]{graphicx}
\usepackage{epstopdf}
\usepackage{braket}
\usepackage[bottom]{footmisc}
\usepackage{bbold}
\usepackage{wasysym}
\usepackage{dcolumn}
\usepackage{bm}
\usepackage{color}
\usepackage{relsize,dsfont,mathrsfs,empheq,verbatim,upgreek}
\usepackage{etoolbox}
\usepackage[caption = false]{subfig}

\newcommand{\reel}{\mathbb{R}}

\newcommand{\diff}{\mathrm{d}}
\newcommand{\e}{\mathrm{e}}
\newcommand{\im}{\mathrm{i}}
\newcommand{\tr}{\operatorname{tr}}

\newcommand{\cD}{\mathcal{D}}

\newcommand{\cW}{\mathcal{W}}

\newcommand{\orho}{\hat{\rho}}
\newcommand{\ovarrho}{\hat{\varrho}}
\newcommand{\osigma}{\hat{\sigma}}
\newcommand{\ogamma}{\hat{\gamma}}
\newcommand{\oGamma}{\hat{\Gamma}}

\newcommand{\oA}{\hat{A}}
\newcommand{\oH}{\hat{H}}
\newcommand{\oh}{\hat{h}}
\newcommand{\oJ}{\hat{J}}
\newcommand{\oL}{\hat{L}}
\newcommand{\ov}{\hat{v}}
\newcommand{\oV}{\hat{V}}
\newcommand{\oU}{\hat{U}}
\newcommand{\oS}{\hat{S}}
\newcommand{\oT}{\hat{T}}

\newcommand{\ovx}{\hat{\boldsymbol{x}}}
\newcommand{\ovp}{\hat{\boldsymbol{p}}}
\newcommand{\ovP}{\hat{\boldsymbol{P}}}

\newcommand{\vp}{\boldsymbol{p}}
\newcommand{\vq}{\boldsymbol{q}}

\newcommand{\vj}{\boldsymbol{j}}

\newcommand{\Boltz}{k_\mathrm{B}}

\definecolor{bitteUnbedingtGegenlesen}{RGB}{255, 255, 0}

\begin{document}
\title{Thermodynamically consistent collisional master equation in a low-density gas with internal structure}

\author{Michael Gaida}%
 \affiliation{
 Naturwissenschaftlich--Technische Fakult\"{a}t, Universit\"{a}t Siegen, 
Walter-Flex-Stra{\ss}e 3, 57068 Siegen, Germany\\
}%

\author{Giulio Gasbarri}
\affiliation{
 Naturwissenschaftlich--Technische Fakult\"{a}t, Universit\"{a}t Siegen, 
Walter-Flex-Stra{\ss}e 3, 57068 Siegen, Germany\\
}%

\author{Stefan Nimmrichter}
\affiliation{
 Naturwissenschaftlich--Technische Fakult\"{a}t, Universit\"{a}t Siegen, 
Walter-Flex-Stra{\ss}e 3, 57068 Siegen, Germany\\
}%


\date{\today}

\begin{abstract}
Quantum thermodynamics with open systems is often based on the quantum optical weak-coupling master equation or on operational repeated interaction models, whereas early works on thermalisation and on decoherence theory were mostly concerned with the kinetics of gas collisions. Here we formulate a master equation for the dynamics of a quantum system under inelastic scattering with a dilute thermal gas in three dimensions, comprised of ancilla particles that also possess internal degrees of freedom. We show thermodynamic consistency when the gas is at thermal equilibrium, irrespective of whether or not the ancillas are in resonance with the system. In contrast, when the internal and the motional state of the gas are thermalised to different temperatures, the gas acts not as two distinct heat baths, but as a structured non-equilibrium reservoir that can generate useful energy through uncontrolled collisions.
\end{abstract}

\maketitle

Understanding how quantum systems in and out of equilibrium exchange energy with their surroundings is a central question in quantum thermodynamics \cite{binder2018thermodynamics}. 
In the weak-coupling limit, one can derive Born-Markov master equations \cite{breuer2002theory} to describe the evolution of a system that continuously couples to a large thermal reservoir consisting of a broad spectrum of field modes. While this perturbative approach and its assumptions are well-justified in quantum optics, they do not generally apply to other physical settings. The original setting studied in classical statistical mechanics, for instance, is that of thermalisation by random discrete collisions, as expressed in Boltzmann’s kinetic theory~\cite{Boltzmann1872} or Einstein’s treatment of Brownian motion~\cite{einstein1905molekularkinetischen}.
Scattering theory is what connects them all and provides a unified view on non-equilibrium processes in open quantum systems.

Quantum models of collisional environments were pioneered by D\"{u}mcke, who rigorously derived the low-density limit for an $N$-level system interacting with an ideal gas~\cite{dumcke1985low}, and by Joos and Zeh, who used single-scattering events to explain the emergence of classicality~\cite{Joos1985}.  
Subsequent work generalized these models to spatial decoherence and to continuous monitoring of motional degrees of freedom~\cite{gallis1990environmental, L.Diósi_1995,hornberger_collisional_2003,schlosshauer2004decoherence, adler_normalization_2006,hornberger_monitoring_2007,vacchini2009quantum, PhysRevA.82.042111}.
While some models do incorporate internal degrees of freedom of the system\cite{PhysRevA.82.042111, Hornberger_2007}, they focus on decoherence and diffusion rather than the full thermodynamics of energy exchange between internal and external degrees of freedom.

An operational perspective is offered by repeated-interaction (collision) models, in which the system undergoes a sequence of unitary collisions with identically prepared ancillas~\cite{ciccarello_quantum_2022}. Such schemes have been used not only to model equilibration, but also to study autonomous quantum machines and information-fueled thermodynamics~\cite{PhysRevX.7.021003,cusumano2024structured}. 
However, physically motivated system-ancilla interactions may drive the system towards a nonequilibrium steady state, and true thermalization is recovered only under careful bath engineering~\cite{prositto2025equilibriumnonequilibriumsteadystates}. 

Recently, Jacob \textit{et al.}~recast collisional dynamics by incorporating a motional degree of freedom in a one-dimensional scattering framework, from which they could obtain  completely positive maps that respect the first and second laws~\cite{jacob_thermalization_2021,jacob_quantum_2022,tabanera2022quantum,tabanera2023thermalization,jacob_universal_2024}. 
They argue that, to ensure thermalisation, it is crucial that the canonical Maxwell–Boltzmann momentum distribution be replaced by a flux-weighted effusion distribution~\cite{ehrich_micro-reversibility_2020}---a consequence of the simplified one-dimensional treatment that, as we will see, one no longer needs in three dimensions.

\begin{figure}
    \centering
    \includegraphics[trim={140 210 100 30},clip, width=1.2\linewidth]{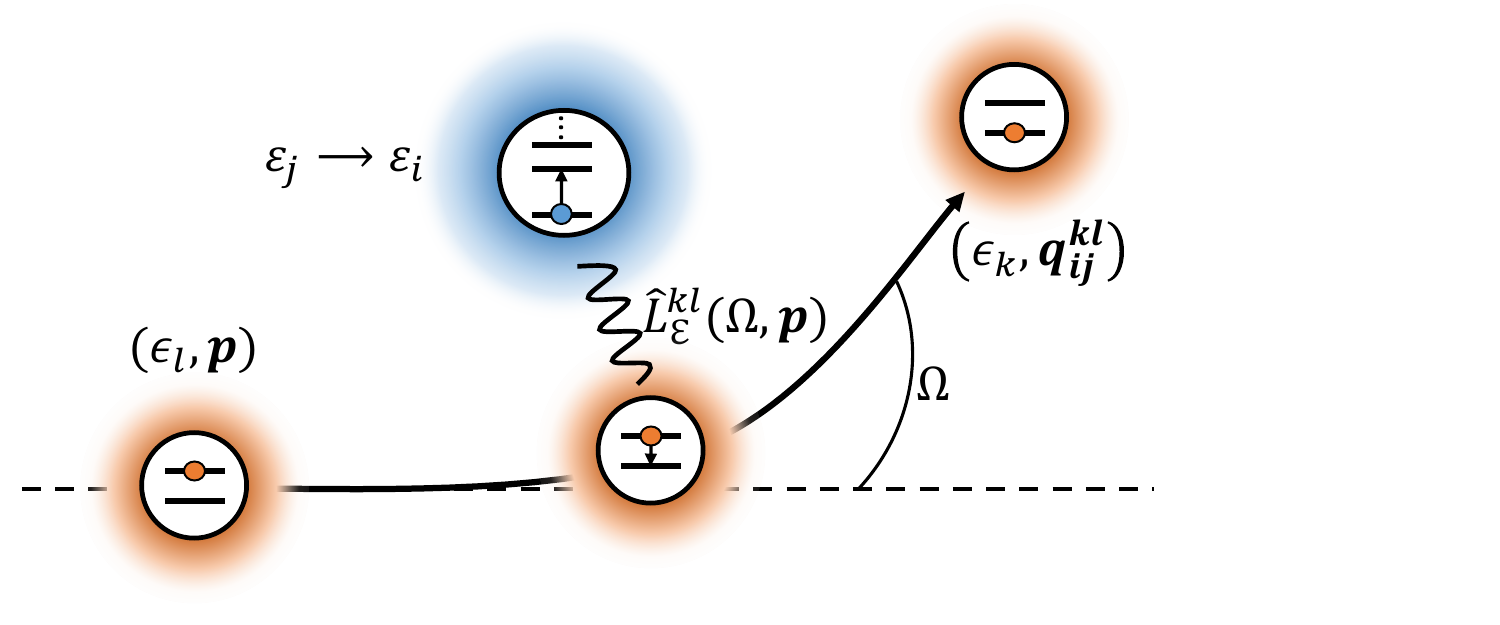}
    \caption{Sketch of an inelastic collision event between a multilevel system and a gas particle with an internal ancilla degree of freedom. Upon entrance into the active scattering region with initial momentum $\vp$, the system-ancilla interaction exchanges system transitions from energy levels $\varepsilon_j$ to $\varepsilon_i$ with ancilla transitions from $\epsilon_l $ to $\epsilon_k$,  changing also the kinetic energy. The particle scatters into the solid angle $\Omega$ with outgoing momentum $\vq_{ij}^{kl}(\Omega,\vp)$. In the master equation \eqref{eq:final_master}, we associate Lindblad operators $\oL_{\mathcal{E}}^{kl}(\Omega,\vp)$ to each such event.}.
    \label{fig:Lindbladoperator}
\end{figure}

Here we formulate a full three-dimensional scattering model for the collisional dynamics of a $d$-level system that exchanges energy with internal degrees of freedom of surrounding gas particles. We derive a master equation that is thermodynamically consistent and achieves proper thermalisation whenever the dilute gas is in thermal equilibrium, regardless of the particles' internal structure. We then demonstrate the possibility of uncontrolled work production in case of a temperature imbalance between the internal and the motional sector of the gas. Crucially, we thus obtain a single non-equilibrium reservoir that is capable of generating ergotropy, but does not resemble two independent heat baths. Unlike similar studies with repeated interaction models \cite{de2018reconciliation, rodrigues2019thermodynamics, ciccarello_quantum_2022}, no precise tuning of interaction times or internal resonances is required.

\paragraph*{Collisional dynamics in a dilute gas---}
Consider a massive quantum system with discrete internal degrees of freedom, characterized by the Hamiltonian $\oH_{\mathrm{S}} = \ovP^2/2M + \oh_{\mathrm{S}}$, which comprises a center-of-mass kinetic energy term and an internal Hamiltonian, $\oh_{\mathrm{S}} = \sum_i \varepsilon_i \ket{\varepsilon_i} \bra{\varepsilon_i}$, with eigenvalues $ \varepsilon_i $ and corresponding eigenvectors $\ket{\varepsilon_i}$. The system interacts with a dilute thermal gas of $ N $ identical, non-interacting particles, each also possessing discrete internal structure that we label the ancilla $A$. The motion shall be confined to a sufficiently large box volume, $ V = L^3 $, such that the continuum limit applies. The Hamiltonian of each gas particle reads, $\oH_{\mathrm{G}} = \ovp^2/2m + \oh_{\mathrm{A}}$, and we describe the ancilla degree of freedom in terms of the eigenvalues and eigenstates of its bare Hamiltonian, $\oh_{\mathrm{A}} = \sum_k \epsilon_k \ket{\epsilon_k} \bra{\epsilon_k}$.
The initial gas state is taken as uncorrelated, $ \oGamma = \ogamma^{\otimes N} $, and each single-particle density operator factorises into an internal and a motional component, $\ogamma = \ogamma_{\mathrm{A}} \otimes \ogamma_{\mathrm{M}}$, with 
\begin{align}
\ogamma_{\mathrm{A}} = \sum_k p(\epsilon_{k}) \ket{\epsilon_{k}}\!\bra{\epsilon_{k}}, \   \ogamma_{\mathrm{M}} = \frac{(2 \pi \hbar)^3}{V} \int d^3\vp \, \mu(\vp ) \ket{\vp }\!\bra{\vp }.
\end{align}
Here, $ p(\epsilon_{k}) $ are (thermal) internal state populations, and $ \mu(\vp ) $ is the Maxwell-Boltzmann momentum distribution at a given inverse temperature $\beta$,
\begin{equation}
\mu(\vp ) = \left( \frac{\beta}{2\pi m} \right)^{3/2} \exp\left( -\frac{\beta \vp^2}{2m} \right),
\end{equation}
normalized to $ \int d^3\vp \, \mu(\vp ) = 1 $. We will abbreviate the joint energy distribution by $\mu_k(\vp ) = p(\epsilon_{k}) \mu(\vp )$.

In the low-density limit, collisions between the system and the gas particles can be regarded as a sequence of statistically independent events~\footnote{Given a characteristic length scale $R$ of the system-gas interaction, the low-density limit demands that $NR^3/V \ll 1$.}. 
We furthermore assume $M\gg m$, allowing us to neglect the system motion and treat it as a static scattering center during collisions.
Each collision is generated by an interaction Hamiltonian between the system and a single gas particle, $\oV =V_0  \ov \otimes g(|\ovx|) $, with $\ovx$ the particle position relative to the scattering center, $g(r)$ a localised function representing the active region around the scatterer, and $\ov$ describing the system-ancilla coupling in this region.
On coarse-grained timescales $ \Delta t $ longer than the duration of the collision event yet short compared to the intrinsic system dynamics, we can describe the net change of the combined state of system and gas particle by the application of the scattering operator, $\Delta \ovarrho_{\rm SG} (t) = \oS \ovarrho_{\rm SG} (t) \oS^\dagger - \ovarrho_{\rm SG} (t-\Delta t)$. Given the Hamiltonian $\oV_\mathrm{I} (t)$ in the interaction picture with respect to the free Hamiltonian $\oH_\mathrm{S} + \oH_{\rm G}$, the scattering operator is formally defined as the time-ordered exponential, 
\begin{equation}\label{eq:S}
    \oS = \mathcal{T} \exp\left[ -\frac{i}{\hbar} \int_{-\infty}^{\infty} dt\, \oV_\mathrm{I}(t) \right] =: \mathds{1} + i\oT ,
\end{equation}
with $\oT$ the transition operator\cite{taylor2012scattering}; see Appendix~\ref{App:Basics}. Expanding the free evolution to first order in $\Delta t$ and using that $\oS$ is unitary, $\oT^\dagger\oT= -i(\oT-\oT^\dagger)$, we obtain
\begin{align}
\Delta \ovarrho_\mathrm{SG}(t) &= -\frac{i}{\hbar}[\oh_\mathrm{S} + \oH_{\rm G}, \ovarrho_{\rm SG}(t)] \Delta t + \frac{i}{2}[\hat{T} + \hat{T}^\dagger, \ovarrho_{\rm SG}(t)] \nonumber\\
&\quad + \left( \hat{T} \ovarrho_{\rm SG}(t) \hat{T}^\dagger - \frac{1}{2}\{ \hat{T}^\dagger \hat{T}, \ovarrho_{\rm SG}(t) \} \right),
\end{align}

Crucially, this expansion is justified only when used to describe observables that vary negligibly under the Hamiltonian during $\Delta t$. However, these observables may still change significantly due to gas collisions.

Summing over the independent collisions with all identically prepared gas particles and tracing them out results in the coarse-grained change of the system state, 
\begin{align}
\label{eq:StartDerivation}
\frac{\Delta \orho_\mathrm{S}}{\Delta t} &= -\frac{i}{\hbar}[\oh_\mathrm{S}, \orho_\mathrm{S}] + \frac{N}{\Delta t} \tr_\mathrm{G} \left\{ \frac{i}{2}[\hat{T} + \hat{T}^\dagger, \orho_\mathrm{S} \otimes \ogamma] \right. \nonumber\\
&\quad \left. +  \left( \hat{T}(\orho_\mathrm{S} \otimes \ogamma) \hat{T}^\dagger -\frac{1}{2} \{ \hat{T}^\dagger \hat{T}, \orho_\mathrm{S} \otimes \ogamma \} \right) \right\}.
\end{align}
The trace is best carried out in the basis of energy and momentum states. After a lengthy calculation detailed in Appendix~\ref{Sec:Master_e}, we are left with a Markovian master equation for the reduced system state, 
\begin{align}\label{eq:final_master}
\partial_t \orho_\mathrm{S} &= -\frac{i}{\hbar}[\hat{h}_\mathrm{S} + \hat{h}_\mathrm{eff}, \orho_\mathrm{S}] + \mathcal{C}\orho_\mathrm{S}, \quad \text{with}\\
\mathcal{C}\orho_\mathrm{S} &= \sum_{\mathcal{E},k,l} \int \diff^3\vp \ \mu_{l}(\vp)    |\vj(\vp)| \int\! \diff \Omega \, \mathcal{D}[\oL_\mathcal{E}^{k,l}( \Omega,\vp)] \orho_\mathrm{S}, \nonumber \\
\hat{h}_\mathrm{eff}  
 &=  -\frac{\pi\hbar^2 n}{m}\sum_{k}\int d^3\vp \, \mu_{k}(\vp ) \hat{L}_0^{kk}(0,\vp ) + \mathrm{h.c.} \nonumber 
\end{align}
Here, we abbreviate $\mathcal{D}[\oL](\orho) = \oL\orho \oL^{\dagger} - \frac{1}{2}\{\oL^{\dagger}\oL,\orho\}$ and $n=N/V$, $\vj(\vp) =  n \vp/m$ denotes the incoming particle flux density, and the solid angle $\Omega$ represents the outgoing direction of a scattered particle, $\vp \to \vq_{i,j}^{k,l} (\Omega)$. The scattering event can thereby induce an incoherent system transition, $\ket{\varepsilon_j} \to \ket{\varepsilon_i}$ of energy $\mathcal{E} = \varepsilon_i - \varepsilon_j$ (or pure dephasing, $\mathcal{E}=0$), accompanied by a transformation of the ancilla state, $\ket{\epsilon_l} \to \ket{\epsilon_k}$, such that $|\vq_{i,j}^{k,l}| = \sqrt{|\vp|^2 - 2m(\mathcal{E}+\epsilon_k-\epsilon_l)}$ by virtue of energy conservation. The respective Lindblad operators read as
\begin{equation}
    \label{eq:FullLind}
    \oL_{\mathcal{E}}^{k,l}( \Omega, \vp)   = \!\!\!\! \sum_{\substack{i,j\\ \varepsilon_{i}-\varepsilon_{j}=\mathcal{E}}} \!\! \sqrt{\frac{|\vq_{i,j}^{k,l}|}{|\vp|}}\, f_{i,j}^{k,l}\left(\vq_{i,j}^{k,l} (\Omega), \vp \right) \ket{\varepsilon_{i}}\!\bra{\varepsilon_{j}},
\end{equation} 
wherein the scattering amplitudes are given by the ``on-shell'' matrix elements of the transition operator \cite{taylor2012scattering},
\begin{equation}
\label{eq:f}
\bra{\varepsilon_i,\epsilon_k, \vq} \oT \ket{\varepsilon_j,\epsilon_l,\vp} = \frac{\delta\left( |\vq_{i,j}^{k,l}|^2 - |\vq|^2 \right)}{\pi \hbar} f_{i,j}^{k,l}(\vq, \vp) .
\end{equation}
The elastic forward scattering amplitudes ($l=k$, $\mathcal{E}=0$, $\Omega=0$) are responsible for the scattering-induced correction of the system Hamiltonian in the third line of \eqref{eq:final_master}, which commutes with the system Hamiltonian.

We remark that the flux-weighted integral in the dissipator $\mathcal{C}$ arises naturally in three dimensions, and it also applies to non-isotropic momentum distributions $\mu(\vp)$ describing, e.g., a directed stream of gas particles. This corroborates  previous results based on one-dimensional motion \cite{jacob_thermalization_2021,jacob_quantum_2022}.

\paragraph*{Thermodynamic Consistency---}
We proceed to show that, under the assumption of micro-reversibility, the scattering master equation \eqref{eq:final_master} respects the laws of thermodynamics. Micro-reversibility holds in the absence of time-dependent interactions or external driving fields, magnetic fields, or any other mechanism that breaks time-reversal symmetry. In terms of the scattering amplitudes \eqref{eq:f}, the symmetry can be stated as
\begin{equation}\label{eq:microreversibility_f}
f_{i,j}^{k,l}(\vq,\vp) = f_{j,i}^{l,k}(-\vp,-\vq).
\end{equation}

The zeroth law requires that the system relax to a Gibbs state when the gas is in thermal equilibrium at an inverse reservoir temperature $\beta = 1/\Boltz T$; that is, when the internal ancilla states $\ogamma_{\rm A}$ and the motional states $\ogamma_{\rm M}$ are Gibbs states of the same $\beta$. 
To verify the zeroth law, we restrict to energy-diagonal system states, 
\begin{equation} \label{eq:ansstate}
\orho_\mathrm{S}(t) = \sum_{i} P(\varepsilon_i,t)\, \ket{\varepsilon_i}\bra{\varepsilon_i}.
\end{equation}
Under the master equation \eqref{eq:final_master}, the energy populations redistribute according to classical rate equations, 
\begin{equation} \label{eq:Therm:ClassMaster}
\dot{P}(\varepsilon_i,t) = \sum_j \left[ R_{i,j}\, P(\varepsilon_j,t) - R_{j,i}\, P(\varepsilon_i,t) \right];
\end{equation}
whereas any off-diagonal matrix element would decohere. The transition rates are given by
\begin{equation}
R_{i,j} = \sum_{k,l} \int \!\!  \diff^3 \vp \, \mu_l(\vp) \frac{n|\vq_{i,j}^{k,l}|}{m} \!\! \int \diff\Omega\, |f_{i,j}^{k,l}(\vq_{i,j}^{k,l}(\Omega),\vp)|^{2} ,
\end{equation}
as derived in Appendix~\ref{App:Rates}. One can then show using \eqref{eq:microreversibility_f} that the rates obey the \textit{local detailed balance} condition,
\begin{equation} \label{eq:Therm:LDB}
\frac{R_{i,j}}{R_{j,i}} = \exp [-\beta (\varepsilon_i - \varepsilon_j)].
\end{equation}
Hence the Gibbs state, $\ogamma_{\rm S} = \exp (-\beta \oh_{\rm S})/Z_{\rm S}$, is a stationary solution of the master equation. 
A violation of micro-reversibility will in general break local detailed balance; systems may still oscillate towards thermal equilibrium, as recently discussed in~\cite{blum2024thermalization}.

The first law follows from the general property of the scattering operator $\oS$ that it commutes with the free Hamiltonian of scatterer and gas particle \cite{taylor2012scattering}, $[\oS, \oh_\mathrm{S} + \oh_\mathrm{A} + \oh_\mathrm{M}] = 0$. The total energy of the system, the ancilla, and the motional degree of freedom ($\oh_\mathrm{M} = \ovp^2/2m$) is thus conserved in each collision event. Given the energy change in each sector, 
\begin{equation}
\Delta E_{X} = \tr \left\{ \oh_X \left( \oS \ovarrho_{\rm SG} \oS^\dagger - \ovarrho_{\rm SG} \right) \right\}, 
\end{equation}
we have $\Delta E_\mathrm{S} + \Delta E_\mathrm{A} + \Delta E_\mathrm{M} = 0$. Since each gas particle represents the thermal environment of the system, we can identify the negative change of ancilla and motional energy as heat that the system receives upon a collision, $\Delta Q = -\Delta E_\mathrm{A} - \Delta E_\mathrm{M}$. No external work is performed, and so the first law can be stated as $\Delta E_\mathrm{S} = \Delta Q$ per collision.
On the level of the master equation \eqref{eq:final_master}, which averages over randomly occurring collision events in time, we consistently identify the heat power as the rate of change of the free system energy,
\begin{equation}\label{eq:1stlaw_ME}
\dot{E}_{\mathrm{S}}(t) = \partial_{t} \tr \left\{ \oh_{\mathrm{S}} \orho_\mathrm{S}(t) \right\} = \tr \left\{ \oh_{\mathrm{S}} \, \mathcal{C}\orho_{\mathrm{S}}(t) \right\} = \dot{Q}(t),
\end{equation}
excluding the collision-induced correction $\oh_{\mathrm{eff}}$.

With the definition of heat power \eqref{eq:1stlaw_ME} at hand, the second law takes the form of a dynamical Clausius inequality for the rate of change of the von Neumann entropy, 
\begin{equation} \label{eq:SecondLaw}
\dot{\mathcal{S}} (t) = - \partial_t \tr \left\{ \orho_\mathrm{S} (t) \ln \orho_\mathrm{S} (t) \right\} \geq \beta \dot{Q}(t).
\end{equation}
This well-known result follows from the monotonicity of the quantum relative entropy under the master equation and the fact that the Gibbs state $\ogamma_\mathrm{S}$ is stationary \cite{spohn2007irreversible,Alicki_1979}.

Consider now a non-equilibrium scenario in which the ancilla and the motional sector of the gas are thermalised at different temperatures, $T_\mathrm{A} \neq T_\mathrm{M}$. This breaks local detailed balance \eqref{eq:Therm:LDB} and enables the production of thermodynamic resources such as work. Since total energy is still conserved in each collision, one may interpret $\Delta E_\mathrm{A}$ and $\Delta E_\mathrm{M}$ as heat flowing into two heat baths at $T_\mathrm{A}$ and $T_\mathrm{M}$, the sum of which matches the energy $-\Delta E_\mathrm{S}$ leaving the system. However, since each collision involves a tripartite interaction between system, ancilla, and motion that allows for direct energy exchange between the latter two parts, one must not mistake $\Delta E_\mathrm{A}$ and $\Delta E_\mathrm{M}$ as heat exchanged separately between the system and two baths. The gas acts as a single structured non-equilibrium reservoir.

\paragraph*{Spin case study---}
We illustrate our framework with a model in which a $(2J+1)$-level system of resonance frequency $\omega_{\rm S}$ interacts with a dilute gas of two-level particles of detuned resonance, $\omega_{\rm A} = \omega_{\rm S} + \Delta$. System operators are represented in terms of the dimensionless spin components $\oJ_x,\oJ_y,\oJ_z$ at fixed total spin quantum number $J$ \cite{sakurai2020modern}, while the ancilla operators are given by Pauli matrices. In particular, $\oh_\mathrm{S} = \hbar \omega_\mathrm{S} \oJ_z$ and $\oh_\mathrm{A} = \hbar \omega_\mathrm{A} \osigma_z$. 
For the interaction Hamiltonian, we let the system and ancilla exchange excitations in a Gaussian active region of size $R$, 
\begin{equation}
    \oV = V_0\, (\oJ_+ \otimes \osigma_- + \oJ_- \otimes \osigma_+) \otimes \exp \left( -\frac{\ovx^2}{2R^2}\right)
\end{equation}
where $\oJ_{\pm} = \oJ_x \pm i\oJ_y$ and $\osigma_{\pm} = (\osigma_x \pm i\osigma_y)/2$ denote the respective raising and lowering operators. Assuming thermal states of ancilla and motion and invoking the Born approximation for the scattering operator, the master equation \eqref{eq:final_master} reduces to a familiar form,
\begin{align}
    \label{eq:ME:Spin}
    \dot{\orho}_\mathrm{S} = -i \omega_\mathrm{S} [\oJ_z, \orho_\mathrm{S}] 
    + \Gamma_+ \cD[\oJ_+] \orho_\mathrm{S} + \Gamma_- \cD[\oJ_-] \orho_\mathrm{S};
\end{align}
see Appendix~\ref{App:MeSpin}. It describes spontaneous excitations and de-excitations across the spin ladder at the rates  
\begin{equation}\label{eq:spinRates}
    \Gamma_{\pm} = \Tilde{\Gamma} \frac{I \left(E_R /\Boltz T_\mathrm{M}, \mp \hbar\Delta/ E_R \right)}{1 + \exp\left( \pm \hbar \omega_\mathrm{A}/\Boltz T_\mathrm{A} \right)},
\end{equation}
with $\Tilde{\Gamma} = n \lambda_{\rm th}^3 V_0^2/(2 \hbar E_R)$, $\lambda_{\rm th} = \sqrt{2\pi\hbar^2/ m \Boltz T_{\rm M}}$ the thermal wavelength of the motional state, $E_R = \hbar^2/2mR^2$, and
\begin{equation} \label{eq:I}
    I(\alpha,s) = \e^s \int_{\max\{0,s\}}^\infty \!\!\!\!\!\!\diff z \ \e^{-(2+\alpha) z} \sinh \left[ 2 \sqrt{z}\sqrt{z-s} \right].
\end{equation}
The rates \eqref{eq:spinRates} obey local detailed balance, $\Gamma_+ = \e^{- \hbar\omega_{\rm S}/\Boltz T}\Gamma_-$, at the effective temperature 
\begin{align}
    \label{eq:Therm:EffTemp}
    T = \frac{\omega_{\rm S}T_\mathrm{A} T_\mathrm{M}}{\omega_{\rm A} T_\mathrm{M} - \Delta T_\mathrm{A}}.
\end{align}
Let us first discuss the case of a gas in thermal equilibrium, $T_\mathrm{A} = T_\mathrm{M} = T$. According to \eqref{eq:ME:Spin}, the system spin will then thermalise regardless of the system-ancilla detuning $\Delta$ or the interaction parameters $E_R,V_0$. In contrast, a repeated interaction model based on excitation exchange with detuned ancillas would make the system relax toward a Gibbs state with a detuning-dependent temperature $\omega_\mathrm{S}T/\omega_\mathrm{A}$ \cite{shu2019almost,prositto2025equilibrium}. This mismatch exemplifies that repeated interaction models must be carefully tuned to the system resonances in order to avoid thermodynamic inconsistency. The scattering master equation, on the other hand, is always consistent, and the detuning merely affects the thermalisation rate, here via the function \eqref{eq:I}. Large detunings will result in strongly suppressed rates; see Appendix~\ref{App:MeSpin} for examples. 

\paragraph*{Non-equilibrium work production---}
The inelastic scattering framework naturally accommodates scenarios in which the combined internal and motional state of the gas particles acts as a structured non-equilibrium reservoir for the system. The simplest such scenario is the above mentioned two-temperature configuration. In experiments with cold atom or molecules, such a configuration could occur in a transient regime, when translational and internal relaxation processes would take place on different timescales or via separate  mechanisms.

Our spin model demonstrates that a non-equilibrium gas can result in a spontaneous buildup of thermodynamic resources through random, uncontrolled collisions. Namely, the effective temperature \eqref{eq:Therm:EffTemp} of the Gibbs state the system equilibrates toward becomes negative when the motional temperature is colder than the internal temperature and the ancilla is sufficiently blue-detuned to the system resonance ($\Delta >0$), such that 
\begin{equation}\label{eq:negT_cond}
    T_\mathrm{M} < \frac{\Delta}{\omega_\mathrm{S}+\Delta} T_\mathrm{A}.
\end{equation}
Negative effective temperatures correspond to population-inverted states, which can provide useful, work-like energy to perform thermodynamic tasks. The useful energy content of a quantum state is commonly measured by the ergotropy~\cite{Allahverdyan_2004}. It is defined as the maximum amount of  energy that one can extract on average from a quantum state by applying a (cyclic) unitary control operation that leaves the system Hamiltonian invariant. For energy-diagonal states \eqref{eq:ansstate}, the ergotropy quantifies the degree of population inversion, 
\begin{equation}\label{eq:ergotropy}
    \cW (t) = \sum_i \varepsilon_i^{\uparrow} \left[ P(\varepsilon_i^{\uparrow},t)- P^{\downarrow}(\varepsilon_i^{\uparrow},t) \right], 
\end{equation}
where $[\varepsilon_i^{\uparrow}]_i$ denotes the sequence of energy values sorted in ascending order and $[P^{\downarrow}(\varepsilon_i^{\uparrow})]_i$ the populations in descending order. Gibbs states of non-negative temperature always have zero ergotropy.

\begin{figure}
    \centering
    \includegraphics[width=1\linewidth]{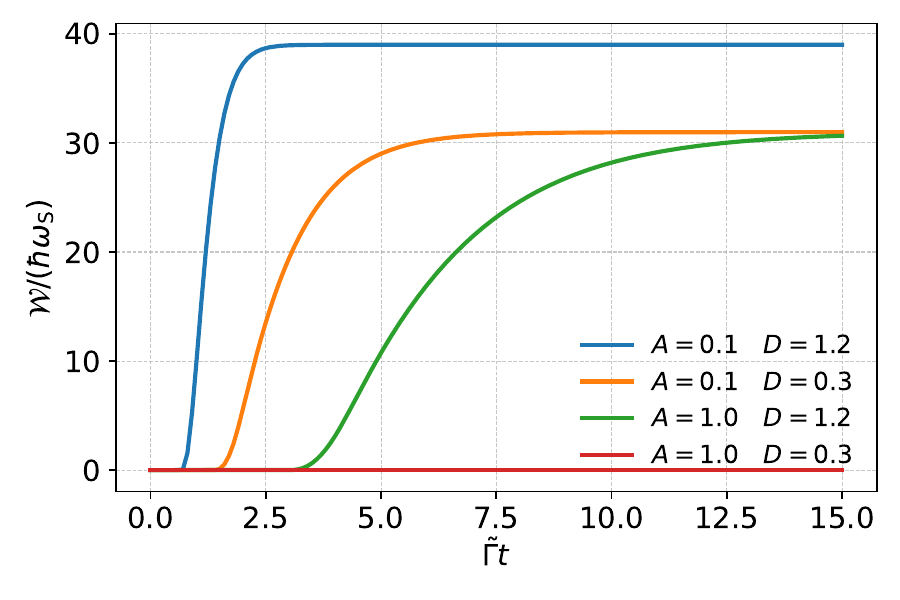}
    \caption{Ergotropy over time for a spin with $J = 20$, initially in its ground state and subjected to a gas of detuned two-level spins with fixed motional temperature, $\Boltz T_\mathrm{M} = E_R$. We evaluate the time evolution \eqref{eq:ME:Spin} for two values each of the detuning and the ancilla temperature, here given as $D=\hbar\Delta/\Boltz T_\mathrm{M}$ and $A=\hbar\omega_\mathrm{A}/\Boltz T_\mathrm{A}$. Ergotropy buildup requires $D>A$. The orange and the green line correspond to the same effective temperature \eqref{eq:Therm:EffTemp}.}
    \label{fig:Therm:Spin:Ergotropy}
\end{figure}

Figure~\ref{fig:Therm:Spin:Ergotropy} shows the time evolution of the ergotropy for a spin system with $J=20$ initially in the ground state, given two exemplary values for the detuning $\Delta$ and the ancilla temperature $T_\mathrm{A}$. One of the four possibilities does not satisfy the condition \eqref{eq:negT_cond} and thus cannot build up any ergotropy.

\paragraph*{Conclusions---}

In this work, we provide a three-dimensional scattering model for the evolution of an open quantum system under spontaneous collisions with a dilute gas of ancilla particles. Contrary to weak coupling models, each collision event may have a strong impact on the system state, but must be sufficiently short and infrequent. Contrary to repeated interaction models, the system-ancilla interaction is mediated and influenced by motional degrees of freedom. We derive a coarse-grained master equation for the system evolution and demonstrate its consistency with the laws of thermodynamics when the gas is at thermal equilibrium, regardless of the precise level structures of system and ancilla. 

Our spin case study demonstrates that, when the internal and the motional state are out of equilibrium, the gas can act as a work reservoir generating population inversion through uncontrolled collisions. Follow-up research should explore possible experiments with quantum gases. In future works, one could also investigate the precise validity regime and non-Markovian effects beyond the low-density assumption.

%



\appendix 
\onecolumngrid

\section{Basics in Scattering theory}
\label{App:Basics}

In this section, we briefly recall the fundamental concepts and notation of scattering theory relevant to our setting. For a comprehensive treatment, we refer the reader to \cite{reed1979iii,taylor2012scattering}.
We consider a two-particle scattering process involving  gas particle and a quantum system with internal degrees of freedom. By transforming to center-of-mass and relative coordinates, the problem reduces to a single effective particle with reduced mass 
$m$ and internal states, moving in an interaction potential.
The total Hamiltonian in the relative frame is given by
\begin{align}
    \hat{H} = \hat{H}_0 + \hat{V},
\end{align}
where $\hat{H}_0$ describes the free evolution of the internal and relative degrees of freedom, and $\hat{V}$ is the interaction potential that depends on the relative coordinates and the internal degrees of freedom of both the system and environment. The free Hamiltonian is taken to be of the form
\begin{equation}
    \oH_0 = \oh + \frac{\ovp^2}{2m} = \sum_\nu \int \diff^3 \vp \left( e_\nu + \frac{\vp^2}{2m} \right) \ket{e_\nu; \vp}\bra{e_\nu; \vp},
\end{equation}
where $m$ is the reduced mass of the two-particle system, $\hat{h}$ is the internal Hamiltonian with eigenstates $\ket{e_\nu}$ and eigenvalues $e_\nu$, and $\hat{\vp }$ is the relative momentum operator. 
The free and interacting dynamics are described by the unitary time-evolution operators $\hat{U}_0(t)$ and $\hat{U}(t)$, generated by the Hamiltonians $\hat{H}_0$ and $\hat{H} = \hat{H}_0 + \hat{V}$, respectively. These operators allow us to define the asymptotic scattering states by comparing the long-time behavior of interacting and non-interacting evolutions.

If the interaction potential $\hat{V}$ is sufficiently localized in space—meaning its position-space matrix elements $\bra{\boldsymbol{x}} \hat{V} \ket{\boldsymbol{x}}$ vanish as $|\boldsymbol{x}| \to \infty$ faster than $1/|\boldsymbol{x}|^{1+\epsilon}$ for some $\epsilon > 0$ - then the influence of the interaction becomes negligible at large distances and long times. 
Under such conditions, the full time evolution asymptotically approaches the free evolution in both the remote past and future. This justifies the definition of asymptotically free incoming and outgoing states and allows for the construction of a well-defined scattering operator $\hat{S}$ as the map between them. These assumptions are satisfied by short-range potentials such as hard-sphere interactions, square wells, or exponentially decaying forces like van der Waals or Yukawa-type interactions~\cite{reed1979iii}. Long-range interactions, such as the Coulomb potential, violate these conditions since their effects persist at large distances. In such cases Standard scattering theory must be modified accordingly. For charged-particle scattering, including Rutherford scattering, Dollard's method~\cite{dollard1964asymptotic} provides an appropriate framework. Here we restrict ourselves to short-range interactions and refer the interested reader to Refs.~\cite{reed1979iii,morchio2016dynamics} for detailed treatments of long-range scattering theory.

Let $\ket{\psi}$ be a state evolving under the full dynamics: $\ket{\psi(t)} = \hat{U}(t) \ket{\psi}$. 
 In the distant past, when the effect of the interaction $\hat{V}$ becomes negligible, the full time evolution effectively reduces to the free evolution generated by $\hat{H}_0$. This motivates the definition of an asymptotically incoming free state $\ket{\psi_{\mathrm{in}}}$, which evolves freely at early times and matches the full evolution in the remote past:
\begin{align}
    \ket{\psi} = \lim_{\tau \to -\infty} \hat{U}^\dagger(\tau)\, \hat{U}_0(\tau)\, \ket{\psi_{\mathrm{in}}}.
\end{align}
Similarly, in the far future, after the scattering has occurred and the interaction again becomes negligible, the system behaves as if it were evolving freely from a corresponding asymptotically outgoing free state $\ket{\psi_{\mathrm{out}}}$:
\begin{align}
    \ket{\psi} = \lim_{\tau \to +\infty} \hat{U}^\dagger(\tau)\, \hat{U}_0(\tau)\, \ket{\psi_{\mathrm{out}}}.
\end{align}
These asymptotic relations imply that the full state $\ket{\psi}$ interpolates between two freely propagating wave packets, with the interaction producing a transition between them. The scattering operator $\hat{S}$ is then defined as the unitary map from incoming to outgoing asymptotic states:
\begin{align}
    \ket{\psi_{\mathrm{out}}} = \hat{S} \ket{\psi_{\mathrm{in}}} = \lim_{\tau \to \infty} \hat{U}_0(-\tau)\, \hat{U}(2\tau)\, \hat{U}_0(-\tau)\, \ket{\psi_{\mathrm{in}}}.
\end{align}
Equivalently the scattering operator can be expressed as a time-ordered exponential involving the interaction-picture potential $
\oV_\mathrm{I}(t) = \oU_{0}^{\dagger}(t) \oV \oU_{0}(t)$:
\begin{align}
    \label{eq:DeffScatt}
    \hat{S} = \lim_{\tau \to \infty} \mathcal{T} \exp\left( -\frac{i}{\hbar} \int_{-\tau}^{\tau} dt\, \hat{V}_\mathrm{I}(t) \right).
\end{align}
This representation highlights $\hat{S}$ as the asymptotic evolution operator in the interaction picture.
It is customary to decompose the scattering operator into the identity operator and the so-called transition operator $\oT$, which encodes the nontrivial part of the scattering process:
\begin{equation}
    \oS = \mathds{1} + \im \oT.
\end{equation}
Unitarity of the scattering operator, $ \oS^{\dagger} \oS =\mathds{1} $, then imposes a constraint on the transition operator.
\begin{align}
\hat{S}^\dagger \hat{S} &= \left( \mathds{1} - i \oT^\dagger \right) \left( \mathds{1} + i \oT \right) = \mathds{1} + i \left( \oT - \oT^\dagger \right) + \oT^\dagger \oT.
\end{align}
Setting this expression equal to the identity leads to the operator version of the optical theorem:
\begin{equation}
\frac{1}{2i} \left( \oT - \oT^\dagger \right) = \frac{1}{2} \oT^\dagger \oT,
\end{equation}
This relation ensures the conservation of probability and expresses the balance between the forward (elastic) and total scattering amplitudes in operator form.
The matrix elements of the transition operator in the energy eigenbasis define the scattering amplitude. 
For incoming and outgoing free states labeled by internal states $\ket{e_{k}}$, $\ket{e_{\nu}}$, and momenta $\vp$, $\vq$ one writes:
\begin{equation}
\label{eq:Basics:TAndAmplitude}
\bra{e_\nu; \vq} \oT \ket{e_\kappa; \vp} = \frac{1}{2\pi \hbar m}  \delta\left( e_\nu + \frac{\vq^2}{2m} - e_\kappa - \frac{\vp^2}{2m} \right)  f_{\nu\kappa}(\vq, \vp).
\end{equation}
Here, the scattering amplitude $f_{\nu \kappa}(\vq,\vp)$, encode information about the probability that the scattering event induces the transition  $\ket{e_{\kappa},\vp}\to \ket{e_{\nu},\vq}$.
The delta function enforces energy conservation during the scattering process. Because of this constraint, the magnitude of the outgoing momentum is fixed by the incoming energy and the internal state transition:
\begin{equation}\label{eq:p_out_magn}
|\vq| =  \sqrt{ \vp^2 - 2m(e_\nu - e_\kappa) } \equiv \vq_{\nu \kappa}.
\end{equation}
In the rest of the section we adopt the notation  $\vq_{\nu \kappa}(\Omega, \vp)$  for the outgoing momentum vector with magnitude defined by the eq.~\eqref{eq:p_out_magn} and direction specified by the solid angle $\Omega$. We will drop the arguments, $\Omega$ and $\vp$, when unambiguous.
An important observable is the differential cross section, which gives the angular distribution of scattered particles:
\begin{equation}
    \frac{\diff \sigma_{\nu \kappa}(\vp)}{\diff \Omega} = \frac{|\vq_{\nu\kappa}|}{|\vp|} |f_{\nu \kappa}(\vq_{\nu\kappa},\vp)|^2.
\end{equation}
Given an incoming particle flux density $\vj(\vp)$, the number of particles $\Delta N_{\nu \kappa}(\Omega, \vp)$ with incoming momentum $\vp$ that get deflected into the solid angle $\Omega$ of size $\Delta \Omega$ and changing the internal state  $\ket{e_\kappa} \to \ket{e_\nu}$ in the time $[t,\Delta t]$ reads,
\begin{equation}
    \Delta N_{\nu \kappa}(\Omega, \vp_\mathrm{in}) = \frac{\diff \sigma_{\nu \kappa}(\vp)}{\diff \Omega} |\boldsymbol{j}| \ \Delta \Omega \ \Delta t.
\end{equation}
An integration over the entire solid angle yields the total cross section, 
\begin{equation}
    \sigma_{\nu\kappa}(\vp) = \int \diff \Omega \ \frac{\diff \sigma_{\nu \kappa}(\vp)}{\diff \Omega}.
\end{equation}

Finally, we discuss the origin of the energy-conserving delta function appearing in eq.~\eqref{eq:Basics:TAndAmplitude}, as it plays a crucial role in motivating the regularization scheme used later in the derivation of the master equation. To illustrate its emergence, we consider the weak-coupling limit, where the interaction term $\oV $ is small and the scattering operator can be approximated to first order by expanding the time-ordered exponential in eq.~\eqref{eq:DeffScatt}. In this regime, the transition operator reads
\begin{equation}
    \hat{T} \approx -\frac{1}{\hbar} \lim_{\tau \to \infty} \int_{-\tau}^\tau dt\, \hat{V}_\mathrm{I}(t)
    = -\frac{1}{\hbar} \lim_{\tau \to \infty} \int_{-\tau}^\tau dt\, e^{i \hat{H}_0 t / \hbar} \hat{V} e^{-i \hat{H}_0 t / \hbar}.
\end{equation}
Evaluating the matrix elements in the energy eigenbasis, we find
\begin{align}
\bra{e_\nu; \vp_\mathrm{out}} \oT \ket{e_\kappa; \vp_\mathrm{in}} 
     &= \lim_{\tau \to \infty} \int_{-\tau}^{+\tau}  \diff t \ e^{\frac{\im}{\hbar} \Delta E t} \bra{\vp_\mathrm{out}, \nu} \oV \ket{\vp_\mathrm{in},\kappa},
\end{align}
where $\Delta E = e_\nu + \frac{\vp_\mathrm{out}^2}{2m} - e_\kappa - \frac{\vp_\mathrm{in}^2}{2m} $ is the total energy difference between the initial and final state. 
The integral over time yields a sinc-type function that reduces to Dirac-delta function in the infinite-time limit:
\begin{equation}
    \delta_\tau(E) = \frac{1}{\pi E} \sin\left( \frac{E \tau}{\hbar} \right).
\end{equation}
Thus, the energy delta function in eq.~\eqref{eq:Basics:TAndAmplitude} arises as the long-time limit ($ \tau \to \infty $) of this regularized function: 
\begin{align}\label{eq:delta_reg}
\delta(E) = \lim_{\tau \to \infty} \delta_\tau(E). 
\end{align}
In the context of master equation derivations, however, the scattering duration is finite, and we must work with the regularized object $ \delta_\tau(E) $, which remains well-defined even at $ E = 0 $:
\begin{equation}\label{eq:delta_reg_0}
    \delta_\tau(0) = \frac{2\tau}{2\pi \hbar}.
\end{equation}
Similar expressions involving the same regularized function $\delta_\tau(E)$ also arise in higher-order terms of the Dyson expansion, where successive time integrals introduce products of energy-conserving sinc functions that encode approximate conservation of energy over finite interaction durations.

\section{Derivation of the Master Equation from Scattering Dynamics}
\label{Sec:Master_e}
This section provides a detailed derivation of eq.~\eqref{eq:final_master} presented in the main text. Given the structure of the gas environment, which includes both motional and internal degrees of freedom, we separate the derivation into two steps. First, we integrate out the motional degrees of freedom, obtaining an effective description for the joint dynamics of the system and internal states of the gas particles. In the second step, we trace out the gas internal states, yielding a reduced dynamical equation for the system alone.
We assume the system--gas correlations relax rapidly between successive collisions, an assumption well-justified in the dilute gas limit where the mean free time between scattering events is large compared to the interaction duration. Accordingly, we model the total state of the system and a single gas particle as a product state,
\begin{align}
\orho_\mathrm{S} \otimes \ogamma_\mathrm{A} \otimes \ogamma_\mathrm{M},
\end{align}
where $\orho_\mathrm{S}$ is the state of the system, $\ogamma_\mathrm{A}$ describes the gas internal state, and $\ogamma_\mathrm{M}$ represents the gas motional state.

\subsection{Step 1: Trace over Motional Degrees of Freedom}
\label{App:TrMotion}

Our first goal is to derive an effective dynamical map governing the joint evolution of the system and the internal degrees of freedom of the gas, under the assumption that the gas is initially in a thermal state.

We begin directly from eq.~\eqref{eq:StartDerivation} of the main text, which expresses the collisional dynamics for the joint system–gas-internal state as:
\begin{align}
\Delta_{c} \ovarrho (t) &= \frac{i}{2}[\hat{T} + \hat{T}^\dagger, \ovarrho (t)] \nonumber\\
&\quad + \left( \hat{T} \ovarrho (t) \hat{T}^\dagger - \frac{1}{2}\{ \hat{T}^\dagger \hat{T}, \ovarrho (t) \} \right),
\end{align}
To isolate the effect of internal transitions, we express the dynamics in the momentum representation and trace over the motional degrees of freedom. For convenience, we group the system $\mathrm{S}$ and the internal component of the gas $\mathrm{A}$ into a single composite subsystem, such that
\begin{equation}
\orho_\mathrm{SA}  \equiv \orho_\mathrm{S} \otimes \ogamma_\mathrm{A},
\end{equation}
Internal states of this composite system are labeled by the Greek letter $ \nu = (i, j) $, denoting a tuple that combines the indices of the internal states of the system and the gas. The corresponding product basis states are written as $ \ket{e_\nu} = \ket{\varepsilon_i} \ket{\epsilon_j} $, where $ \ket{\varepsilon_i} $ and $ \ket{\epsilon_j} $ are energy eigenstates of the system and gas internal Hamiltonians, respectively.

The collisional contribution to the evolution of $\orho_\mathrm{SA}$ is obtained via tracing over the trace over the motional degrees of freedom.
In the momentum basis, the dissipative compontent of the collisional contribution takes the form
\begin{align}
    \label{eq:Therm:FirstLindblad}
    &N\tr_{\mathrm{M}} \left\{  \oT \orho_\mathrm{SA}(t) \otimes \hat{\gamma}_\mathrm{M} \oT^\dagger - \frac{1}{2} \left\{ \oT^\dagger \oT, \orho_\mathrm{SA}(t) \otimes \hat{\gamma}_\mathrm{M} \right\}   \right\} \nonumber \\ 
    &= (2 \pi \hbar)^3 \frac{N}{L^3}
    \int \diff^3 \vq \ \bra{\vq} \left( \oT \orho_\mathrm{SA}(t) \otimes \mu(\ovp) \oT^\dagger - \frac{1}{2} \left\{ \oT^\dagger \oT, \orho_\mathrm{SA}(t) \otimes \mu(\ovp)  \right\}
    \right) \ket{\vq} \nonumber \\   
    &= (2 \pi \hbar)^3 n \int \diff^3 \vp \ \mu(\vp)\int \diff^3 \vq \ \cD\left[ \bra{\vq} \oT \ket{\vp} \right] \orho_\mathrm{SA}, 
\end{align}
where we used $\ogamma_\mathrm{M} = (2\pi\hbar/L)^3 \mu(\vp)$  to express the motional Gibbs state in the momentum basis and identified the gas density $n = N/L^3$. The superoperator $\mathcal{D}[\oA]\orho = \oA \orho \oA^\dagger - \tfrac{1}{2}{\oA^\dagger \oA, \orho}$ enotes the standard Lindblad dissipator.

Our next step is to evaluate the (still operator-valued) matrix elements $\bra{\vq} \oT \ket{\vp}$ in terms of the two-body scattering amplitude $f_{\nu \kappa}(\vq, \vp)$. This will allow us to rewrite the collision term fully in terms of the internal state transitions of the system and gas, as well as the momentum exchange between the incoming and outgoing relative motion. We now proceed with this calculation.
Using the standard relation between the scattering operator and the on-shell scattering amplitude, we write:
\begin{align}
    \label{eq:Therm:OperatorTemp}
    \bra{\vq} \hat{T} \ket{\vp} 
    = \frac{1}{2\pi \hbar m} \sum_{\nu, \kappa} 
    \delta\left( \frac{\vq^2}{2m} + e_\nu - \frac{\vp^2}{2m} - e_\kappa \right)
    f_{\nu \kappa}(\vq, \vp)\, \ket{e_\nu} \bra{e_\kappa},
\end{align}
where $ e_\nu \equiv \varepsilon_i + \epsilon_j $ is the total internal energy associated with the product state $ \ket{e_\nu} = \ket{\varepsilon_i} \ket{\epsilon_j} $ of system and gas internal states.

Formally, Eq.~\eqref{eq:Therm:FirstLindblad} already provides the collisional dynamics in Lindblad form. However, the presence of energy-conserving delta functions in eq.~\eqref{eq:Therm:OperatorTemp} obscures the structure of possible transitions. To clarify this, observe that each term in the superoperator $ \mathcal{D}[\bra{\vq} \hat{T} \ket{\vp}] \ovarrho_\mathrm{SA} $ contains a product of two delta functions:
\begin{align}
    \label{eq:Therm:DeltaIdentity}
    &\delta\left( \frac{\vq^2}{2m} + e_{\nu_2} - \frac{\vp^2}{2m} - e_{\kappa_2} \right)
    \delta\left( \frac{\vq^2}{2m} + e_{\nu_1} - \frac{\vp^2}{2m} - e_{\kappa_1} \right) = \delta\left( \frac{\vq^2}{2m} + e_{\nu_2} - \frac{\vp^2}{2m} - e_{\kappa_2} \right)
    \delta\left( \mathcal{E}_{\nu_1 \kappa_1} - \mathcal{E}_{\nu_2 \kappa_2} \right),
\end{align}
where we define the internal energy differences $ \mathcal{E}_{\nu \kappa} \equiv e_\nu - e_\kappa $. where  to obtain the l.h.s. we used the first delta function to eliminate $ \vq^2/2m $ from the argument of the second.

The second delta function, $ \delta(\mathcal{E}_{\nu_1 \kappa_1} - \mathcal{E}_{\nu_2 \kappa_2}) $, constrains the internal energy differences to be equal. Since the set of internal energies is discrete, the argument of the delta function takes values on a discrete set. As a result we have that:
\begin{align}
\delta(\mathcal{E}_{\nu_1 \kappa_1} - \mathcal{E}_{\nu_2 \kappa_2}) 
= \begin{cases}
\delta(0), & \text{if } \mathcal{E}_{\nu_1 \kappa_1} = \mathcal{E}_{\nu_2 \kappa_2}, \\
0, & \text{otherwise}.
\end{cases}
\end{align}
However, the appearence of the $\delta(0)$ is formally ill-defined and requires regularization. 
As shown in the previous section (see Eq.~\eqref{eq:delta_reg}), this Dirac delta arises as the limiting case of a finite-time energy resolution kernel. In particular, for finite interaction time  $\Delta t $, the delta function over internal energy differences satisfies
\begin{equation}
    \label{eq:Therm:DeltaReg}
    \delta\left( \mathcal{E}_{\nu_1 \kappa_1} - \mathcal{E}_{\nu_2 \kappa_2} \right)
    = \frac{\Delta t}{2\pi \hbar}\, \Tilde{\delta} \left( \mathcal{E}_{\nu_1 \kappa_1} - \mathcal{E}_{\nu_2 \kappa_2} \right),
\end{equation}
where \( \Tilde{\delta} \) denotes a Kronecker delta on the discrete set of internal energy differences.

Finally, we use a known distributional identity to rewrite,
\begin{equation}
    \delta \left( \frac{\vq^2}{2m} + \mathcal{E}_{\nu_1} - \frac{\vp^2}{2m} - \mathcal{E}_{\kappa_1} \right) =\frac{m}{|\vq|} \delta\left( |\vq| - \sqrt{\vp^2 - 2m \mathcal{E}_{\nu_1 \kappa_1}} \right).
\end{equation}
Putting everything together, we obtain,
\begin{equation}
    \delta \left( \frac{\vq^2}{2m} + e_{\nu_2} - \frac{\vp^2}{2m} - e_{\kappa_2} \right) \delta \left( \frac{\vq^2}{2m} + e_{\nu_1} - \frac{\vp^2}{2m} - e_{\kappa_1} \right) = \frac{m}{|\vq|} \delta\left( |\vq| - \sqrt{\vp^2 - 2m \mathcal{E}_{\nu_1 \kappa_1}} \right)  \frac{\Delta t}{2 \pi \hbar} \Tilde{\delta} \left( \mathcal{E}_{\nu_1 \kappa_1} - \mathcal{E}_{\nu_2 \kappa_2} \right).
\end{equation}

With the distributional identity at hand, we can rephrase the first part of the dissipator,
\begin{align}
    &\int \diff^3 \vq \ \bra{\vq} \oT \ket{\vp} \orho_\mathrm{SA} \bra{\vp} \oT^\dagger \ket{\vq} 
    = \frac{1}{(2 \pi \hbar)^3} \frac{\Delta t}{m} \int \diff \Omega \ \sum_{\substack{\nu_1,\kappa_1,\nu_2,\kappa_2 \\ \mathcal{E}_{\nu_1,\kappa_1} = \mathcal{E}_{\nu_2,\kappa_2}}}  \nonumber  \\ 
    &|\vq_{\nu_1 \kappa_1}| \  f_{\nu_1 \kappa_1}\left( \vq_{\nu_1 \kappa_1}, \vp \right) \ket{e_{\nu_1}} \bra{e_{\kappa_1}} \orho_\mathrm{SA} \ket{e_{\kappa_2}} \bra{e_{\nu_2}}  f^*_{\nu_2 \kappa_2}\left( \vq_{\nu_2 \kappa_2}, \vp \right) \nonumber \\ 
    &= \frac{1}{(2 \pi \hbar)^3} \frac{|\vp| \Delta t}{m} \int \diff \Omega \ 
    \sum_{ \mathcal{E} } 
    \left( \sum_{\substack{\nu_1,\kappa_1 \\ \mathcal{E}_{\nu_1, \kappa_1}=\mathcal{E}  }} \sqrt{\frac{|\vq_{\nu_1 \kappa_1}|}{|\vp|}} f_{\nu_1 \kappa_1}\left( \vq_{\nu_1 \kappa_1}, \vp \right) \ket{e_{\nu_1}} \bra{e_{\kappa_1}} \right) \nonumber \\ 
    &\orho_\mathrm{SA}
    \left( \sum_{\substack{\nu_2,\kappa_2 \\ \mathcal{E}_{\nu_2, \kappa_2}=\mathcal{E}  }} \sqrt{\frac{|\vq_{\nu_2 \kappa_2}|}{|\vp|}} f_{\nu_2 \kappa_2}\left( |\vq_{\nu_2 \kappa_2}, \vp \right) \ket{e_{\nu_2}} \bra{e_{\kappa_2}} \right)^\dagger \nonumber \\ 
    &= \frac{1}{(2 \pi \hbar)^3} \frac{|\vp| \Delta t}{m} \int \diff \Omega \ \sum_{\mathcal{E}} \oL_\mathcal{E}(\Omega,\vp) \orho_\mathrm{SA} \oL^\dagger_\mathcal{E}( \Omega,\vp),
\end{align}
where, in the first step, we used the delta identity \eqref{eq:Therm:DeltaIdentity} and integrated over $|\vq|$ in spherical coordinates. Second, we split the condition $\mathcal{E}_{\nu_1,\kappa_1} = \mathcal{E}_{\nu_2,\kappa_2}$ by using
\begin{align}
    &\sum_{\mathcal{E} \in \reel} \delta^{(0)} \left( \mathcal{E}_{\nu_1,\kappa_1} - \mathcal{E}  \right) \ \delta^{(0)} \left( \mathcal{E}_{\nu_2,\kappa_2} - \mathcal{E}  \right) 
    = \delta^{(0)} \left( \mathcal{E}_{\nu_2,\kappa_2} - \mathcal{E}_{\nu_1,\kappa_1}  \right) \sum_{\mathcal{E}l} \delta^{(0)} \left( \mathcal{E}_{\nu_1,\kappa_1} - \mathcal{E}  \right) \nonumber \\ 
    &= \delta^{(0)} \left( \mathcal{E}_{\nu_2,\kappa_2} - \mathcal{E}_{\nu_1,\kappa_1}  \right),
\end{align}
and multiplied and divided the expression by $|\vp|$. Note that $\vq_{ij}^{nk}$ depends on the incoming momentum $\vp$ and the solid angle $\Omega$. For the sake of readability, we will supress this dependence in our derivations. In order to simplify the expression, we introduce the Lindblad operators,
\begin{equation}
   \oL_\mathcal{E}(\Omega, \vp) = \sum_{\substack{\nu,\kappa \\ \mathcal{E}_{\nu \kappa}=\mathcal{E}  }} \sqrt{\frac{|\vq_{\nu \kappa}|}{|\vp|}} f_{\nu \kappa}\left( \vq_{\nu \kappa}, \vp \right) \ket{\varepsilon_\nu} \bra{\varepsilon_\kappa}.
\end{equation}
With these, the first term of the dissipator becomes, 
\begin{equation}
 \int \diff^3 \vq \ \bra{\vq} \oT \ket{\vp} \orho_\mathrm{SA} \bra{\vp} \oT^\dagger \ket{\vq} = \frac{1}{(2 \pi \hbar)^3} \frac{|\vp| \Delta t}{m} \int \diff \Omega \ \sum_{\mathcal{E}} \oL_\mathcal{E}(\Omega,\vp) \orho_\mathrm{SA} \oL^\dagger_\mathcal{E}( \Omega,\vp). 
\end{equation}
This is already sufficient to fully fix the generator of the dynamical semi group described in \eqref{eq:final_master}. Alternatively, one can follow analogous steps to find the remaining part of the dissipator,
\begin{equation}
    \int \diff^3 \vq \  \left\{ \bra{\vq} \oT \ket{\vp}, \orho_{\mathrm{SA}} \right\} = 
    \frac{1}{(2 \pi \hbar)^3} \frac{|\vp| \Delta t}{m} \int \diff \Omega \ \sum_{\mathcal{E}} \left\{ \oL_\mathcal{E}^\dagger(\Omega,\vp) \oL_\mathcal{E}(\Omega,\vp), \orho_{\mathrm{SA}} \right\},
\end{equation}
and the effective Hamiltonian,
\begin{align}
    \oh_\mathrm{eff,SA} &= - \frac{2 \pi \hbar^2}{m} n \int \diff^3 \vp \ \mu(\vp) \sum_{\substack{\nu,\kappa \\ \varepsilon_\nu = \epsilon_\kappa }} \frac{1}{2} \left( f_{\nu,\kappa}(\vp,\vp) + f_{\kappa,\nu}^*(\vp,\vp)  \right) \ket{\varepsilon_\nu}\bra{\varepsilon_\kappa} \\ 
    &= - \frac{2 \pi \hbar^2}{m} n \int \diff^3 \vp \ \mu(\vp) \frac{1}{2} \left( \oL_0(0,\vp) + \oL^\dagger_0(0,\vp) \right). \nonumber
\end{align}

Altogether, the effective evolution of the internal degrees of freedom of system and gas read
\begin{align}
    \frac{\diff}{\diff t} \orho_\mathrm{SA} 
    = -\frac{\im}{\hbar} \left[ \oh_\mathrm{eff}, \orho_\mathrm{SA} \right] 
    +  \int \diff^3 \vp \ \mu(\vp) |\vj(\vp)| \int \diff \Omega \sum_{\mathcal{E}} \cD \left[ \oL_\mathcal{E}(\Omega,\vp) \right] \orho_\mathrm{SA }.
\end{align}

\subsection{Step 2: Trace over the internal degrees of freedom of the gas}
\label{App:TrAncilla}

We proceed by taking the trace over the internal degrees of freedom of the gas. In order to do so, we split the Greek multi indices again, $\nu \to (i,j)$, and denote the gas indices by superscripts,
\begin{align}
    \mathcal{E}_{\nu \kappa} &\to \mathcal{E}_{ij}^{kl}  \\ 
      f_{\nu \kappa}\left( \vq,\vp \right) &\to  f_{ij}^{kl} \left( \vq,\vp \right).
\end{align}
We define the populations of the ancilla Gibbs state to be $p(\epsilon_k) = \bra{\epsilon_k}\gamma_\mathrm{A}\ket{\epsilon_k}$ and use the linearity of the partial trace to focus on a single Lindblad operator,
\begin{align}
    \tr_A \left\{  \oL_\mathcal{E}(\Omega,\vp) \orho_\mathrm{S}  \otimes \ogamma_{\mathrm{A}} \oL^\dagger_\mathcal{E}(\Omega,\vp)  \right\}  
    &= \sum_{n,k} p(\epsilon_k) \bra{\epsilon_n} \oL_\mathcal{E}(\Omega,\vp) \ket{\epsilon_k} \orho_\mathrm{S} \bra{\epsilon_k} \oL^\dagger_\mathcal{E}(\Omega,\vp) \ket{\epsilon_n}
\end{align}
where we expanded $\ogamma_\mathrm{A}$ in the energy eigenbasis and wrote out the partial trace in this basis.

Further, we can simplify the energy conservation condition $\mathcal{E}^{n,k}_{i,j}=\mathcal{E}$ under the $\sum_\mathcal{E}$ sum because the same indices $n,k$ appear in both Lindblad operators: For initial $\orho_\mathrm{S} = \ket{\varepsilon_i} \bra{\varepsilon_i^\prime}$, we find
\begin{align}
    &\bra{\varepsilon_j}\left( \sum_{\mathcal{E}} \bra{\epsilon_n} \oL_\mathcal{E} (\Omega,\vp) \ket{\epsilon_k} \orho_\mathrm{S}  \bra{\epsilon_k} \oL^\dagger_\mathcal{E} (\Omega,\vp) \ket{\epsilon_n} \right) \ket{\varepsilon_j^\prime}
    \nonumber \\ 
    &\propto \sum_{\mathcal{E}} 
    \delta^{(0)} \left( \mathcal{E}^{n k}_{ji}-\mathcal{E} \right)
    \delta^{(0)} \left( \mathcal{E}^{n k}_{j^\prime i^\prime}- \mathcal{E} \right) \nonumber \\ 
    &= \delta^{(0)} \left( \mathcal{E}_{ji}-\mathcal{E}_{j^\prime  i^\prime} \right) 
    \underbrace{\sum_{\mathcal{E}} \delta^{(0)} \left( \mathcal{E}^{n k}_{j^\prime i^\prime}-\mathcal{E} \right)}_{=1} \nonumber \\ 
    &= \sum_{\mathcal{E}} \delta^{(0)} \left( \mathcal{E}_{j i}-\mathcal{E} \right)
    \delta^{(0)} \left( \mathcal{E}_{j^\prime i^\prime} - \mathcal{E} \right),
\end{align} 
where we replaced $\mathcal{E}$ by $\mathcal{E}^{n,k}_{j^\prime, i^\prime}$ in the first Kronecker delta by the virtue of the second one, noted that the remaining sum yields one, and in the end, we factorized the remaining term by introducing a new sum over energies $\mathcal{E}$. Summarized, the term can be rephrased as,
\begin{equation}
\tr_A \left\{  \oL_\mathcal{E}(\Omega,\vp) \orho_\mathrm{S}  \otimes \ogamma_{\mathrm{A}} \oL^\dagger_\mathcal{E}(\Omega,\vp)  \right\}  
    = \sum_{nk} p(\epsilon_k) \oL_{\mathcal{E}}^{n k}(\Omega, \vp) \orho_\mathrm{S} \oL_{\mathcal{E}}^{n k \dagger}(\Omega, \vp),
\end{equation}
where we introduced the new Lindblad operators,
\begin{equation}
    \oL_{\mathcal{E}}^{n k}(\Omega, \vp) = \sum_{\substack{j i \\ \mathcal{E}_{ij}= \mathcal{E} } } \sqrt{\frac{|\vq_{ij}^{nk}|}{|\vp|}} f_{ij}^{n k}\left( \vq_{ij}^{nk} , \vp \right) \ket{\varepsilon_i}\bra{\varepsilon_j}.
\end{equation}

As before, the generator is already fixed by the first term of the dissipator and we find the full master equation \eqref{eq:final_master} with the effective Hamiltonian 
\begin{align}
    \oh_\mathrm{eff} &= \sum_{k} p(\epsilon_k) \bra{\epsilon_k} \oh_\mathrm{eff,SA} \ket{\epsilon_k} \nonumber \\ 
    &= -\frac{2 \pi \hbar^2}{m} n 
    \sum_{k} \int \diff^3 \vp \  \mu_k(\vp) \sum_{\substack{ji \\ \varepsilon_{j} = \varepsilon_{i} }} \frac{1}{2} \left( f_{ij}^{k k}(\vp, \vp) + f_{ji}^{kk *}(\vp, \vp)  \right) \ket{\varepsilon_i} \bra{\varepsilon_j}, \nonumber \\ 
    &= -\frac{2 \pi \hbar^2}{m} n \sum_{k} \int \diff^3 \vp \  \mu_k(\vp) \frac{1}{2} \left( \oL_0^{kk}(0,\vp) + \mathrm{h.c.} \right),
\end{align}
where we introduced $\mu_k(\vp) = p(\epsilon_k) \mu(\vp)$.

\section{Rate equation and local detailed balance}
\label{App:Rates}

We derive a classical master equation for the system populations and show that the rates fulfill the local detailed balance condition. The assumption of a non-degenerate Hamiltonian simplifies the calculation because $\varepsilon_{i} = \varepsilon_{j}$ is equivalent to $i=j$. Due to this fact, the effective Hamiltonian becomes diagonal
\begin{equation}
     \oh_{\mathrm{eff}} = -\frac{2 \pi \hbar^2}{m} n \sum_{k} p(\epsilon_k) \int \diff^3 \vp \ \mu(\vp) \sum_{i} \frac{1}{2} \left(  f_{ii}^{kk}(\vp,\vp) + f_{ii}^{kk*}(\vp,\vp) \right) \ket{\varepsilon_i} \bra{\varepsilon_i} ,
\end{equation}
which only shifts the existing energy levels but does not introduce transitions between them. 
Similar conclusions can be drawn for the dissipator, where direct calculation yields
\begin{align}
    \bra{\varepsilon_j} \mathcal{D}\left[ \oL_\mathcal{E}^{nk} \right] (\ket{\varepsilon_i}\bra{\varepsilon_i}) \ket{\varepsilon_j^\prime} &\propto \delta_{j j^\prime} \\ 
    \bra{\varepsilon_j} \mathcal{D}\left[ \oL_\mathcal{E}^{nk} \right] (\ket{\varepsilon_i}\bra{\varepsilon_i^\prime}) \ket{\varepsilon_j} &\propto \delta_{i i^\prime}.
\end{align}
Hence, transitions $\ket{\varepsilon_i}\bra{\varepsilon_i} \longleftrightarrow \ket{\varepsilon_j}\bra{\varepsilon_j^\prime}$ occur if and only if $j=j^\prime$. Conclusively, the diagonal elements evolve independently from the non-diagonal elements and vice versa.

Now, assuming a diagonal state $\orho_\mathrm{S}(t) = \sum_{i} p_{i}(t) \ket{\varepsilon_i}\bra{\varepsilon_i}$ we find the classical master equation 
\begin{equation}
    \dot{P}(\varepsilon_i, t) = \bra{\varepsilon_i} \mathcal{C}(\orho_\mathrm{S}) \ket{\varepsilon_i} 
    = \sum_{j} \bra{\varepsilon_i} \mathcal{C}(\ket{\varepsilon_j} \bra{\varepsilon_j}) \ket{\varepsilon_i} \ P(\varepsilon_j, t).
\end{equation}
Note that the free time evolution has no impact on the populations since $\orho_\mathrm{S}$ commutes with $\oh_\mathrm{S}$ and $\oh_\mathrm{eff}$. Written out, the transition term reads,
\begin{align}
    \bra{\varepsilon_i} \mathcal{C}(\ket{\varepsilon_j} \bra{\varepsilon_j}) \ket{\varepsilon_i} = \sum_{n,k} p(\epsilon_k) \int \diff^3 \vp \ \mu(\vp) |\vj(\vp)| \int \diff \Omega \ \sum_{\mathcal{E}} \bra{\varepsilon_i} \mathcal{D} \left[ \oL_\mathcal{E}^{nk}(\Omega, \vp) \right] \left( \ket{\varepsilon_j}\bra{\varepsilon_j} \right) \ket{\varepsilon_i}. \nonumber
\end{align} 
The individual terms read,
\begin{align}
    &\sum_{\mathcal{E}} \bra{\varepsilon_i} \mathcal{D} \left[ \oL_\mathcal{E}^{nk}(\Omega, \vp) \right] \left( \ket{\varepsilon_j}\bra{\varepsilon_j} \right) \ket{\varepsilon_i} \nonumber \\
    &= \sum_{\mathcal{E}} \left( \left| \bra{\varepsilon_i} \oL_\mathcal{E}^{nk}(\Omega, \vp) \ket{\varepsilon_j} \right|^2 - \delta_{ij} \sum_{l} \left| \bra{\varepsilon_l} \oL_\mathcal{E}^{nk}(\Omega, \vp) \ket{\varepsilon_j} \right|^2  \right) \nonumber \\
    &= \frac{\diff \sigma_{ij}^{nk}(\Omega,\vp)}{\diff \Omega} - \delta_{ij} \sum_l \frac{\diff \sigma_{ij}^{nk}(\Omega,\vp)}{\diff \Omega} \nonumber \\ 
    &= \frac{|\vq_{ij}^{nk}|}{|\vp|} f_{ij}^{nk}\left(\vq_{ij}^{nk}, \vp \right) - \delta_{ij} \sum_l \frac{|\vq_{lj}^{nk}|}{|\vp|} f_{lj}^{nk}\left(\vq_{lj}^{nk}, \vp \right).
\end{align}
Due to the linearity of integration, we can integrate term by term and define the rates $R_{ij}$,
\begin{align}
    \label{eq:Therm:Rate}
   &\sum_{n,k} \int \diff^3 \vp \ \mu_k(\vp) |\vj(\vp)| \int \diff \Omega \ \frac{|\vq_{ij}^{nk}|}{|\vp|} \left\vert f_{ij}^{nk}\left(\vq_{ij}^{nk}, \vp \right) \right\vert^2  \nonumber \\ 
   &= \sum_{n,k} \int \diff^3 \vp \ \mu_k(\vp) \int \diff \Omega \ \frac{n|\vq_{ij}^{nk}|}{m} \left\vert f_{ij}^{nk}\left(\vq_{ij}^{nk}, \vp \right) \right\vert^2 \equiv R_{ij}.
\end{align}

We find the well-known structure of a Markov chain,
\begin{equation}
    \label{eq:Therm:ClassMaster}
    \dot{P}(\varepsilon_i,t) =\sum_{j} \bra{\varepsilon_i} \mathcal{C}(\ket{\varepsilon_j} \bra{\varepsilon_j}) \ket{\varepsilon_i} P(\varepsilon_j,t) = \sum_{j} \left( R_{ij} P(\varepsilon_j, t) - R_{ji} P(\varepsilon_i, t) \right)  , 
\end{equation} 
where the first term describes the flow of probability from all states $j$ into the state of consideration $i$ and the second term describes the total flow out of state $i$.

Under the assumptions of time reversal symmetry, i.e. $f_{ij}^{nk}(\vq,\vp) = f_{ji}^{kn}(-\vp, -\vq)$, we now show that the rates $R_{ij}$ obey local detailed balance. To this end, we expand the energy conservation condition, on which the relation between ingoing and outgoing momentum is based, again by a delta function,
\begin{align}
    \label{eq:Therm:LDBDerivaton}
    R_{ij}
    &= n \sum_{n k} 
    \int \diff^3 \vq \int \diff^3 \vp \ \mu_k(\vp) \delta 
    \left( \frac{\vq^2}{2m} - \left[ \frac{\vp^2}{2m} -\mathcal{E}_{ij}^{nk} \right] \right) |f_{ij}^{nk} (\vq,\vp)|^2 \nonumber \\ 
    &= n \sum_{n_\mathrm{A} k_\mathrm{A}} 
    \int \diff^3 \vq \int \diff^3 \vp \ \mu_k(\vp) \delta
    \left( \frac{\vp^2}{2m} - \left[ \frac{\vq^2}{2m} -\mathcal{E}_{j i}^{k n} \right] \right) |f_{ji}^{kn} (\vp,\vq)|^2 \nonumber \\ 
    &= n \sum_{n k} \e^{\beta \mathcal{E}^{n k}} p(\epsilon_n) \int \diff^3 \vq \int \diff^3 \vp \ \mu(\vq)  \e^{\beta \mathcal{E}_{j i}^{k,n} } \nonumber \\ & \quad \delta
    \left( \frac{\vp^2}{2m} - \left[ \frac{\vq^2}{2m} - \mathcal{E}  _{j,i}^{k,n} \right] \right) |f_{j i}^{k n} (\vp,\vq)|^2 \nonumber \\ 
    &= \e^{\beta \mathcal{E}_{j i}} R_{ji}.
\end{align}
In the first step, we used micro-reversibility. Since $\vp$ and $\vq$ are integrated over all directions, the extra signs can be  changed back. In the same step, we also used $\mathcal{E}_{ij}^{nk} = -\mathcal{E}_{ji}^{kn}$ to rewrite the delta function. Second, we used $p(\epsilon_k)/p(\epsilon_n) = \e^{\beta \mathcal{E}^{n k}}$ and the delta function to replace the argument of the Gaussian
\begin{equation}
   \mu(\vp) \delta 
    \left( \frac{\vp^2}{2m} - \left[ \frac{\vq^2}{2m} -\mathcal{E}_{ji}^{kn} \right] \right) 
    =  \e^{\beta \mathcal{E}_{ji}^{kn}} \mu(\vq) \delta 
    \left( \frac{\vp^2}{2m} - \left[ \frac{\vq^2}{2m} -\mathcal{E}_{ji}^{kn} \right] \right).  \nonumber
\end{equation}
In the third step, we noted $\e^{\beta \mathcal{E}^{nk}}  \e^{\beta \mathcal{E}_{ji}^{kn}} = \e^{\beta \mathcal{E}_{ji}}$ and recognized the formula for the rate in the opposite direction. This shows that the classical master equation for the populations of a mixed state fulfills the local detailed balance condition, 
\begin{equation} 
\label{eq:AppTherm:LDB}
\frac{R_{jk}}{R_{kj}} = e^{-\beta (\epsilon_j - \epsilon_k)}.
\end{equation}

If we assume our system to be in the thermal state $\ogamma_\mathrm{S} = \exp(-\beta \oh_\mathrm{S})/Z$, which has the populations $P(\varepsilon_j) \propto \e^{-\beta \varepsilon_j}$,
the evolution in this state according to \eqref{eq:Therm:ClassMaster} is 
\begin{equation}
    \dot{P}(\varepsilon_i,t) \propto \sum_j \underbrace{ \left( R_{ij} \e^{-\beta \varepsilon_j} - R_{ji} \e^{-\beta \epsilon_i} \right)}_{=0}  = 0,
\end{equation}
where, due to \eqref{eq:AppTherm:LDB},  the in-and out-flow of probability cancels for each pair of energy levels individually. Hence, the populations of the Gibbs state do not change under the master equation. On operator level, this reads
\begin{equation}
    \mathcal{C}\ogamma_\mathrm{S} = 0.
\end{equation} 

\section{Derivation of the master equation for a spin}
\label{App:MeSpin}

In this appendix, we derive the master equation for the $j$-spin in a qubit gas with Gaussian interaction. We differentiate between the temperature of the motional $T_\mathrm{M}$ and the internal $T_\mathrm{A}$ degree of feedom of the gas. The scattering amplitude for a Gaussian shaped interaction, 
\begin{equation}
    \oV = V_0 \ov \otimes \exp \left( - \frac{\ovx^2}{2R^2} \right),
\end{equation}
in first order Born approximation read, 
\begin{align}
    f_{ij}^{nk} \left( \vp_2, \vp_1 \right) 
    &= -(2\pi)^2 \hbar m  \bra{\varepsilon_i, \epsilon_n; \vp_2} \oV \ket{\varepsilon_j, \epsilon_k; \vp_1} \nonumber \\ 
    &= -\sqrt{\frac{\pi}{2}} R \frac{V_0}{E_R} \bra{\varepsilon_i, \epsilon_n} \ov \ket{\varepsilon_j, \epsilon_k} \exp \left( \frac{(\vp_2 - \vp_1)^2}{2p_R^2} \right).
\end{align}
We introduced the momentum scale $p_R = \hbar/R$ and the energy scale $E_R = p_R^2/(2m)$ of the potential. Plugged into the Lindblad operator, the first term in the Lindblad equation for $\orho_\mathrm{S} = \ket{\varepsilon_j} \bra{\varepsilon_j^\prime}$ can be rephrased, 
\begin{align}
    &\sum_\mathcal{E} \int \diff^3 \vp \ \mu(\vp) |\vj(\vp)| \int \diff \Omega \ \oL_\mathcal{E}^{nk}(\Omega,\vp) \orho_\mathrm{S} \oL_\mathcal{E}^{nk \dagger}(\Omega,\vp) \\ 
    &= \frac{\pi}{2} R^2 \left( \frac{V_0}{E_R} \right)^2 \sum_{\substack{ii^\prime \\ \mathcal{E}_{ij} = \mathcal{E}_{i^\prime j^\prime}}} 
    v_{ij}^{nk} \left( v_{i^\prime j^\prime}^{nk} \right)^* 
    \int \diff^3 \vp \ \mu(\vp) |\vj(\vp)| \int \diff \Omega \
     \sqrt{\frac{|\vq_{ij}^{nk}|}{|\vp|}} \exp \left( - \frac{(\vq_{ij}^{nk} - \vp)^2}{2p_R^2} \right) \ket{\varepsilon_i}\bra{\varepsilon_i^\prime} \nonumber \\ 
    &= \frac{\pi}{2} R^2 \left( \frac{V_0}{E_R} \right)^2 \sum_{\substack{ii^\prime \\ \mathcal{E}_{ij} = \mathcal{E}_{i^\prime j^\prime}}} 
    v_{ij}^{nk} \left( v_{i^\prime j^\prime}^{nk} \right)^* \left( \frac{2 \sqrt{2 \pi} n p_R^2}{p_\mathrm{th}^3} E_R \ 
    I\left( \frac{E_R}{\Boltz T_\mathrm{M}}, \frac{\mathcal{E}_{ij}^{nk}}{E_R} \right) \right) \ket{\varepsilon_i}\bra{\varepsilon_i^\prime} \nonumber \\
    &= \Tilde{\Gamma}
    \sum_\mathcal{E} \left( \sum_{\substack{ij \\ \mathcal{E}_{ij} = \mathcal{E}}} \sqrt{I\left( \frac{E_R}{\Boltz T_\mathrm{M}}, \frac{\mathcal{E}_{ij}^{nk}}{E_R} \right)} v_{ij}^{nk} \ket{\varepsilon_i}\bra{\varepsilon_j}  \right) \orho_\mathrm{S} \left( \sum_{\substack{i^\prime j^\prime \\ \mathcal{E}_{i^\prime j^\prime} = \mathcal{E}}} \sqrt{I\left( \frac{E_R}{\Boltz T_\mathrm{M}}, \frac{\mathcal{E}_{ij}^{nk}}{E_R} \right)} v_{ij}^{nk} \ket{\varepsilon_i^\prime}\bra{\varepsilon_j^\prime}  \right)^\dagger \nonumber 
\end{align}
where solving the $\Omega$ integral in spherical coordinates that have $\vp$ as the polar axis, allows a reduction to the one-dimensional integral $I(\alpha,s)$. Combining the prefactors into new rate $\Tilde{\Gamma}$, yields a reduced Lindblad equation. The integral and effective rate read,
\begin{align}
    \Tilde{\Gamma} &= \frac{n \lambda_\mathrm{th}^3 V_0^2}{2 \hbar E_R} \\ 
    I(\alpha,s) &= \e^s \int_{\max\{0,s\}}^\infty \diff z \ \e^{-(2+\alpha) z} \sinh \left( 2 \sqrt{z}\sqrt{z-s} \right).
\end{align}

Recall that the interaction between spin and gas is characterized by the free Hamiltonians for system, $\oh_\mathrm{S} = \hbar \omega_\mathrm{S} \oJ_z$, and ancilla, $\oh_\mathrm{A} = \hbar \omega_\mathrm{A} \osigma_z$ and the interaction Hamiltonian,
\begin{equation}
    \ov = \oJ_+ \otimes \osigma_- + \oJ_- \otimes \osigma_+.
\end{equation}
Since the changes of angular momentum introduced to the system by $\ov$ are limited to $\pm \hbar$, the possible energy transitions are $\mathcal{E}_{i,j} \in \left\{ \pm \hbar \omega_\mathrm{S}\right\}$. Since any excitation of the system comes with a de-excitation of the ancilla, the total change of internal energy is restricted to $\mathcal{E}_{ij}^{nk} \in \left\{ \pm \hbar \Delta \right\}$ and we only have to consider two Lindblad operators, one for $n=0$,$k=1$ and one for $n=1$, $k=0$ .

\begin{figure}
    \centering
    \includegraphics[width=0.65\linewidth]{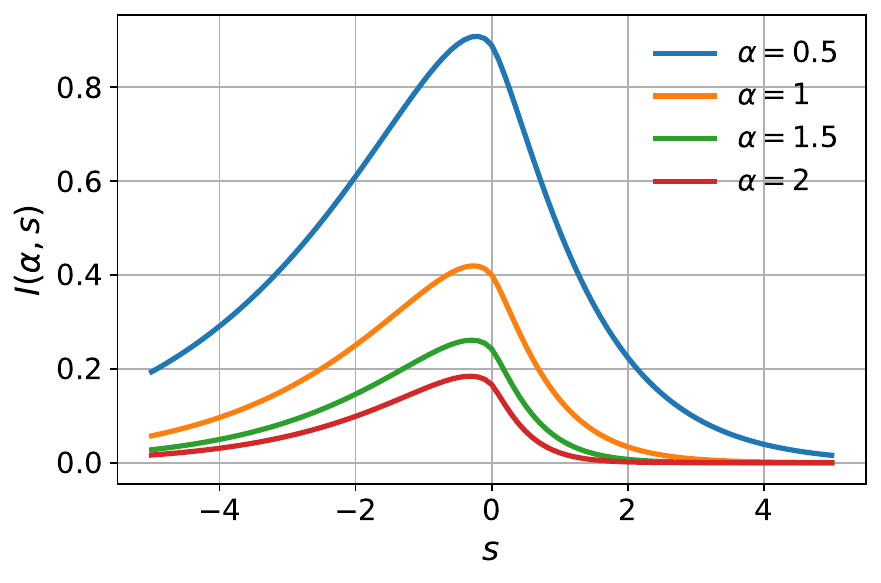}
    \caption{Numerically calculated values of the parameter integral $I(\alpha,s)$. The value of the integral is strongly suppressed for large $\alpha$ as well as large $|s|$. It decays exponentially faster for positve $s$ than for negative $s$. }
    \label{Fig:Integral}
\end{figure}

Altogether, the dissipator can be written as,
\begin{align}
    \mathcal{C} \left[ \orho_\mathrm{S} \right] 
    &= \Tilde{\Gamma}  p(\epsilon_1) I\left( \frac{E_R}{\Boltz T}, \frac{-\hbar \Delta }{E_R}  \right)  \cD \left[ \oJ_+ \right] + \Tilde{\Gamma} p(\epsilon_0) I\left( \frac{E_R}{\Boltz T}, \frac{+\hbar \Delta }{E_R}  \right)  \cD \left[ \oJ_- \right] \nonumber \\ 
    &=  \Gamma_+ \cD \left[ \oJ_+ \right] + \Gamma_- \cD \left[ \oJ_- \right],
\end{align}
with the rates,
\begin{equation}
    \Gamma_\pm = \Tilde{\Gamma} \frac{I\left( E_R/(\Boltz T_\mathrm{M}), \mp \hbar \Delta/E_R  \right)}{1+\exp\left( \pm \frac{\hbar \omega_\mathrm{A}}{\Boltz T_\mathrm{A}} \right)}.
\end{equation}
Note that the rates go to zero for large detunings, i.e. $\hbar |\Delta| \gg E_R$, as can be seen from Fig.~\ref{Fig:Integral}.

\begin{thebibliography}{39}%
\makeatletter
\providecommand \@ifxundefined [1]{%
 \@ifx{#1\undefined}
}%
\providecommand \@ifnum [1]{%
 \ifnum #1\expandafter \@firstoftwo
 \else \expandafter \@secondoftwo
 \fi
}%
\providecommand \@ifx [1]{%
 \ifx #1\expandafter \@firstoftwo
 \else \expandafter \@secondoftwo
 \fi
}%
\providecommand \natexlab [1]{#1}%
\providecommand \enquote  [1]{``#1''}%
\providecommand \bibnamefont  [1]{#1}%
\providecommand \bibfnamefont [1]{#1}%
\providecommand \citenamefont [1]{#1}%
\providecommand \href@noop [0]{\@secondoftwo}%
\providecommand \href [0]{\begingroup \@sanitize@url \@href}%
\providecommand \@href[1]{\@@startlink{#1}\@@href}%
\providecommand \@@href[1]{\endgroup#1\@@endlink}%
\providecommand \@sanitize@url [0]{\catcode `\\12\catcode `\$12\catcode `\&12\catcode `\#12\catcode `\^12\catcode `\_12\catcode `\%12\relax}%
\providecommand \@@startlink[1]{}%
\providecommand \@@endlink[0]{}%
\providecommand \url  [0]{\begingroup\@sanitize@url \@url }%
\providecommand \@url [1]{\endgroup\@href {#1}{\urlprefix }}%
\providecommand \urlprefix  [0]{URL }%
\providecommand \Eprint [0]{\href }%
\providecommand \doibase [0]{https://doi.org/}%
\providecommand \selectlanguage [0]{\@gobble}%
\providecommand \bibinfo  [0]{\@secondoftwo}%
\providecommand \bibfield  [0]{\@secondoftwo}%
\providecommand \translation [1]{[#1]}%
\providecommand \BibitemOpen [0]{}%
\providecommand \bibitemStop [0]{}%
\providecommand \bibitemNoStop [0]{.\EOS\space}%
\providecommand \EOS [0]{\spacefactor3000\relax}%
\providecommand \BibitemShut  [1]{\csname bibitem#1\endcsname}%
\let\auto@bib@innerbib\@empty
\bibitem [{\citenamefont {Binder}\ \emph {et~al.}(2018)\citenamefont {Binder}, \citenamefont {Correa}, \citenamefont {Gogolin}, \citenamefont {Anders},\ and\ \citenamefont {Adesso}}]{binder2018thermodynamics}%
  \BibitemOpen
  \bibfield  {author} {\bibinfo {author} {\bibfnamefont {F.}~\bibnamefont {Binder}}, \bibinfo {author} {\bibfnamefont {L.~A.}\ \bibnamefont {Correa}}, \bibinfo {author} {\bibfnamefont {C.}~\bibnamefont {Gogolin}}, \bibinfo {author} {\bibfnamefont {J.}~\bibnamefont {Anders}},\ and\ \bibinfo {author} {\bibfnamefont {G.}~\bibnamefont {Adesso}},\ }\href@noop {} {\bibfield  {journal} {\bibinfo  {journal} {Fundamental Theories of Physics}\ }\textbf {\bibinfo {volume} {195}} (\bibinfo {year} {2018})}\BibitemShut {NoStop}%
\bibitem [{\citenamefont {Breuer}\ and\ \citenamefont {Petruccione}(2002)}]{breuer2002theory}%
  \BibitemOpen
  \bibfield  {author} {\bibinfo {author} {\bibfnamefont {H.-P.}\ \bibnamefont {Breuer}}\ and\ \bibinfo {author} {\bibfnamefont {F.}~\bibnamefont {Petruccione}},\ }\href@noop {} {\emph {\bibinfo {title} {The theory of open quantum systems}}}\ (\bibinfo  {publisher} {OUP Oxford},\ \bibinfo {year} {2002})\BibitemShut {NoStop}%
\bibitem [{\citenamefont {Boltzmann}(1872)}]{Boltzmann1872}%
  \BibitemOpen
  \bibfield  {author} {\bibinfo {author} {\bibfnamefont {L.}~\bibnamefont {Boltzmann}},\ }\href@noop {} {\bibfield  {journal} {\bibinfo  {journal} {Sitzungsberichte der Kaiserlichen Akademie der Wissenschaften}\ }\textbf {\bibinfo {volume} {66}},\ \bibinfo {pages} {275} (\bibinfo {year} {1872})}\BibitemShut {NoStop}%
\bibitem [{\citenamefont {Einstein}(1905)}]{einstein1905molekularkinetischen}%
  \BibitemOpen
  \bibfield  {author} {\bibinfo {author} {\bibfnamefont {A.}~\bibnamefont {Einstein}},\ }\href@noop {} {\bibfield  {journal} {\bibinfo  {journal} {Annalen der physik}\ }\textbf {\bibinfo {volume} {4}} (\bibinfo {year} {1905})}\BibitemShut {NoStop}%
\bibitem [{\citenamefont {D{\"u}mcke}(1985)}]{dumcke1985low}%
  \BibitemOpen
  \bibfield  {author} {\bibinfo {author} {\bibfnamefont {R.}~\bibnamefont {D{\"u}mcke}},\ }\href@noop {} {\bibfield  {journal} {\bibinfo  {journal} {Commun. Math. Phys.}\ }\textbf {\bibinfo {volume} {97}},\ \bibinfo {pages} {331} (\bibinfo {year} {1985})}\BibitemShut {NoStop}%
\bibitem [{\citenamefont {Joos}\ and\ \citenamefont {Zeh}(1985)}]{Joos1985}%
  \BibitemOpen
  \bibfield  {author} {\bibinfo {author} {\bibfnamefont {E.}~\bibnamefont {Joos}}\ and\ \bibinfo {author} {\bibfnamefont {H.~D.}\ \bibnamefont {Zeh}},\ }\href {https://doi.org/10.1007/BF01725541} {\bibfield  {journal} {\bibinfo  {journal} {Z. Phys. B Condens. Matter}\ }\textbf {\bibinfo {volume} {59}},\ \bibinfo {pages} {223} (\bibinfo {year} {1985})}\BibitemShut {NoStop}%
\bibitem [{\citenamefont {Gallis}\ and\ \citenamefont {Fleming}(1990)}]{gallis1990environmental}%
  \BibitemOpen
  \bibfield  {author} {\bibinfo {author} {\bibfnamefont {M.~R.}\ \bibnamefont {Gallis}}\ and\ \bibinfo {author} {\bibfnamefont {G.~N.}\ \bibnamefont {Fleming}},\ }\href@noop {} {\bibfield  {journal} {\bibinfo  {journal} {Phys. Rev. A}\ }\textbf {\bibinfo {volume} {42}},\ \bibinfo {pages} {38} (\bibinfo {year} {1990})}\BibitemShut {NoStop}%
\bibitem [{\citenamefont {Diósi}(1995)}]{L.Diósi_1995}%
  \BibitemOpen
  \bibfield  {author} {\bibinfo {author} {\bibfnamefont {L.}~\bibnamefont {Diósi}},\ }\href {https://doi.org/10.1209/0295-5075/30/2/001} {\bibfield  {journal} {\bibinfo  {journal} {EPL}\ }\textbf {\bibinfo {volume} {30}},\ \bibinfo {pages} {63} (\bibinfo {year} {1995})}\BibitemShut {NoStop}%
\bibitem [{\citenamefont {Hornberger}\ and\ \citenamefont {Sipe}(2003)}]{hornberger_collisional_2003}%
  \BibitemOpen
  \bibfield  {author} {\bibinfo {author} {\bibfnamefont {K.}~\bibnamefont {Hornberger}}\ and\ \bibinfo {author} {\bibfnamefont {J.~E.}\ \bibnamefont {Sipe}},\ }\href {https://doi.org/10.1103/PhysRevA.68.012105} {\bibfield  {journal} {\bibinfo  {journal} {Phys. Rev. A}\ }\textbf {\bibinfo {volume} {68}},\ \bibinfo {pages} {012105} (\bibinfo {year} {2003})}\BibitemShut {NoStop}%
\bibitem [{\citenamefont {Schlosshauer}(2004)}]{schlosshauer2004decoherence}%
  \BibitemOpen
  \bibfield  {author} {\bibinfo {author} {\bibfnamefont {M.}~\bibnamefont {Schlosshauer}},\ }\href@noop {} {\bibfield  {journal} {\bibinfo  {journal} {Reviews of Modern physics}\ }\textbf {\bibinfo {volume} {76}},\ \bibinfo {pages} {1267} (\bibinfo {year} {2004})}\BibitemShut {NoStop}%
\bibitem [{\citenamefont {Adler}(2006)}]{adler_normalization_2006}%
  \BibitemOpen
  \bibfield  {author} {\bibinfo {author} {\bibfnamefont {S.~L.}\ \bibnamefont {Adler}},\ }\href {https://doi.org/10.1088/0305-4470/39/45/015} {\bibfield  {journal} {\bibinfo  {journal} {J. Phys. A-Math.}\ }\textbf {\bibinfo {volume} {39}},\ \bibinfo {pages} {14067} (\bibinfo {year} {2006})}\BibitemShut {NoStop}%
\bibitem [{\citenamefont {Hornberger}(2007{\natexlab{a}})}]{hornberger_monitoring_2007}%
  \BibitemOpen
  \bibfield  {author} {\bibinfo {author} {\bibfnamefont {K.}~\bibnamefont {Hornberger}},\ }\href {https://doi.org/10.1209/0295-5075/77/50007} {\bibfield  {journal} {\bibinfo  {journal} {EPL}\ }\textbf {\bibinfo {volume} {77}},\ \bibinfo {pages} {50007} (\bibinfo {year} {2007}{\natexlab{a}})}\BibitemShut {NoStop}%
\bibitem [{\citenamefont {Vacchini}\ and\ \citenamefont {Hornberger}(2009)}]{vacchini2009quantum}%
  \BibitemOpen
  \bibfield  {author} {\bibinfo {author} {\bibfnamefont {B.}~\bibnamefont {Vacchini}}\ and\ \bibinfo {author} {\bibfnamefont {K.}~\bibnamefont {Hornberger}},\ }\href@noop {} {\bibfield  {journal} {\bibinfo  {journal} {Phys. Rep.}\ }\textbf {\bibinfo {volume} {478}},\ \bibinfo {pages} {71} (\bibinfo {year} {2009})}\BibitemShut {NoStop}%
\bibitem [{\citenamefont {Smirne}\ and\ \citenamefont {Vacchini}(2010)}]{PhysRevA.82.042111}%
  \BibitemOpen
  \bibfield  {author} {\bibinfo {author} {\bibfnamefont {A.}~\bibnamefont {Smirne}}\ and\ \bibinfo {author} {\bibfnamefont {B.}~\bibnamefont {Vacchini}},\ }\href {https://doi.org/10.1103/PhysRevA.82.042111} {\bibfield  {journal} {\bibinfo  {journal} {Phys. Rev. A}\ }\textbf {\bibinfo {volume} {82}},\ \bibinfo {pages} {042111} (\bibinfo {year} {2010})}\BibitemShut {NoStop}%
\bibitem [{\citenamefont {Hornberger}(2007{\natexlab{b}})}]{Hornberger_2007}%
  \BibitemOpen
  \bibfield  {author} {\bibinfo {author} {\bibfnamefont {K.}~\bibnamefont {Hornberger}},\ }\href {https://doi.org/10.1209/0295-5075/77/50007} {\bibfield  {journal} {\bibinfo  {journal} {Europhysics Letters}\ }\textbf {\bibinfo {volume} {77}},\ \bibinfo {pages} {50007} (\bibinfo {year} {2007}{\natexlab{b}})}\BibitemShut {NoStop}%
\bibitem [{\citenamefont {Ciccarello}\ \emph {et~al.}(2022)\citenamefont {Ciccarello}, \citenamefont {Lorenzo}, \citenamefont {Giovannetti},\ and\ \citenamefont {Palma}}]{ciccarello_quantum_2022}%
  \BibitemOpen
  \bibfield  {author} {\bibinfo {author} {\bibfnamefont {F.}~\bibnamefont {Ciccarello}}, \bibinfo {author} {\bibfnamefont {S.}~\bibnamefont {Lorenzo}}, \bibinfo {author} {\bibfnamefont {V.}~\bibnamefont {Giovannetti}},\ and\ \bibinfo {author} {\bibfnamefont {G.~M.}\ \bibnamefont {Palma}},\ }\href@noop {} {\bibfield  {journal} {\bibinfo  {journal} {Phys. Rep.}\ }\textbf {\bibinfo {volume} {954}},\ \bibinfo {pages} {1} (\bibinfo {year} {2022})}\BibitemShut {NoStop}%
\bibitem [{\citenamefont {Strasberg}\ \emph {et~al.}(2017)\citenamefont {Strasberg}, \citenamefont {Schaller}, \citenamefont {Brandes},\ and\ \citenamefont {Esposito}}]{PhysRevX.7.021003}%
  \BibitemOpen
  \bibfield  {author} {\bibinfo {author} {\bibfnamefont {P.}~\bibnamefont {Strasberg}}, \bibinfo {author} {\bibfnamefont {G.}~\bibnamefont {Schaller}}, \bibinfo {author} {\bibfnamefont {T.}~\bibnamefont {Brandes}},\ and\ \bibinfo {author} {\bibfnamefont {M.}~\bibnamefont {Esposito}},\ }\href {https://doi.org/10.1103/PhysRevX.7.021003} {\bibfield  {journal} {\bibinfo  {journal} {Phys. Rev. X}\ }\textbf {\bibinfo {volume} {7}},\ \bibinfo {pages} {021003} (\bibinfo {year} {2017})}\BibitemShut {NoStop}%
\bibitem [{\citenamefont {Cusumano}\ and\ \citenamefont {De~Chiara}(2024)}]{cusumano2024structured}%
  \BibitemOpen
  \bibfield  {author} {\bibinfo {author} {\bibfnamefont {S.}~\bibnamefont {Cusumano}}\ and\ \bibinfo {author} {\bibfnamefont {G.}~\bibnamefont {De~Chiara}},\ }\href@noop {} {\bibfield  {journal} {\bibinfo  {journal} {New J. Phys}\ }\textbf {\bibinfo {volume} {26}},\ \bibinfo {pages} {023001} (\bibinfo {year} {2024})}\BibitemShut {NoStop}%
\bibitem [{\citenamefont {Prositto}\ \emph {et~al.}(2025{\natexlab{a}})\citenamefont {Prositto}, \citenamefont {Forbes},\ and\ \citenamefont {Segal}}]{prositto2025equilibriumnonequilibriumsteadystates}%
  \BibitemOpen
  \bibfield  {author} {\bibinfo {author} {\bibfnamefont {A.}~\bibnamefont {Prositto}}, \bibinfo {author} {\bibfnamefont {M.}~\bibnamefont {Forbes}},\ and\ \bibinfo {author} {\bibfnamefont {D.}~\bibnamefont {Segal}},\ }\href {https://arxiv.org/abs/2501.05392} {\bibinfo {title} {Equilibrium and nonequilibrium steady states with the repeated interaction protocol: Relaxation dynamics and energetic cost}} (\bibinfo {year} {2025}{\natexlab{a}}),\ \Eprint {https://arxiv.org/abs/2501.05392} {arXiv:2501.05392 [quant-ph]} \BibitemShut {NoStop}%
\bibitem [{\citenamefont {Jacob}\ \emph {et~al.}(2021)\citenamefont {Jacob}, \citenamefont {Esposito}, \citenamefont {Parrondo},\ and\ \citenamefont {Barra}}]{jacob_thermalization_2021}%
  \BibitemOpen
  \bibfield  {author} {\bibinfo {author} {\bibfnamefont {S.~L.}\ \bibnamefont {Jacob}}, \bibinfo {author} {\bibfnamefont {M.}~\bibnamefont {Esposito}}, \bibinfo {author} {\bibfnamefont {J.~M.}\ \bibnamefont {Parrondo}},\ and\ \bibinfo {author} {\bibfnamefont {F.}~\bibnamefont {Barra}},\ }\href {https://doi.org/10.1103/PRXQuantum.2.020312} {\bibfield  {journal} {\bibinfo  {journal} {PRX Quantum}\ }\textbf {\bibinfo {volume} {2}},\ \bibinfo {pages} {020312} (\bibinfo {year} {2021})}\BibitemShut {NoStop}%
\bibitem [{\citenamefont {Jacob}\ \emph {et~al.}(2022)\citenamefont {Jacob}, \citenamefont {Esposito}, \citenamefont {Parrondo},\ and\ \citenamefont {Barra}}]{jacob_quantum_2022}%
  \BibitemOpen
  \bibfield  {author} {\bibinfo {author} {\bibfnamefont {S.~L.}\ \bibnamefont {Jacob}}, \bibinfo {author} {\bibfnamefont {M.}~\bibnamefont {Esposito}}, \bibinfo {author} {\bibfnamefont {J.~M.~R.}\ \bibnamefont {Parrondo}},\ and\ \bibinfo {author} {\bibfnamefont {F.}~\bibnamefont {Barra}},\ }\href {https://doi.org/10.22331/q-2022-06-29-750} {\bibfield  {journal} {\bibinfo  {journal} {Quantum}\ }\textbf {\bibinfo {volume} {6}},\ \bibinfo {pages} {750} (\bibinfo {year} {2022})}\BibitemShut {NoStop}%
\bibitem [{\citenamefont {Tabanera}\ \emph {et~al.}(2022)\citenamefont {Tabanera}, \citenamefont {Luque}, \citenamefont {Jacob}, \citenamefont {Esposito}, \citenamefont {Barra},\ and\ \citenamefont {Parrondo}}]{tabanera2022quantum}%
  \BibitemOpen
  \bibfield  {author} {\bibinfo {author} {\bibfnamefont {J.}~\bibnamefont {Tabanera}}, \bibinfo {author} {\bibfnamefont {I.}~\bibnamefont {Luque}}, \bibinfo {author} {\bibfnamefont {S.~L.}\ \bibnamefont {Jacob}}, \bibinfo {author} {\bibfnamefont {M.}~\bibnamefont {Esposito}}, \bibinfo {author} {\bibfnamefont {F.}~\bibnamefont {Barra}},\ and\ \bibinfo {author} {\bibfnamefont {J.~M.}\ \bibnamefont {Parrondo}},\ }\href@noop {} {\bibfield  {journal} {\bibinfo  {journal} {NJP}\ }\textbf {\bibinfo {volume} {24}},\ \bibinfo {pages} {023018} (\bibinfo {year} {2022})}\BibitemShut {NoStop}%
\bibitem [{\citenamefont {Tabanera-Bravo}\ \emph {et~al.}(2023)\citenamefont {Tabanera-Bravo}, \citenamefont {Parrondo}, \citenamefont {Esposito},\ and\ \citenamefont {Barra}}]{tabanera2023thermalization}%
  \BibitemOpen
  \bibfield  {author} {\bibinfo {author} {\bibfnamefont {J.}~\bibnamefont {Tabanera-Bravo}}, \bibinfo {author} {\bibfnamefont {J.~M.}\ \bibnamefont {Parrondo}}, \bibinfo {author} {\bibfnamefont {M.}~\bibnamefont {Esposito}},\ and\ \bibinfo {author} {\bibfnamefont {F.}~\bibnamefont {Barra}},\ }\href@noop {} {\bibfield  {journal} {\bibinfo  {journal} {PRL}\ }\textbf {\bibinfo {volume} {130}},\ \bibinfo {pages} {200402} (\bibinfo {year} {2023})}\BibitemShut {NoStop}%
\bibitem [{\citenamefont {Jacob}\ \emph {et~al.}(2024)\citenamefont {Jacob}, \citenamefont {Goold}, \citenamefont {Landi},\ and\ \citenamefont {Barra}}]{jacob_universal_2024}%
  \BibitemOpen
  \bibfield  {author} {\bibinfo {author} {\bibfnamefont {S.~L.}\ \bibnamefont {Jacob}}, \bibinfo {author} {\bibfnamefont {J.}~\bibnamefont {Goold}}, \bibinfo {author} {\bibfnamefont {G.~T.}\ \bibnamefont {Landi}},\ and\ \bibinfo {author} {\bibfnamefont {F.}~\bibnamefont {Barra}},\ }\href {https://doi.org/10.1103/PhysRevLett.133.207101} {\bibfield  {journal} {\bibinfo  {journal} {Phys. Rev. Lett.}\ }\textbf {\bibinfo {volume} {133}},\ \bibinfo {pages} {207101} (\bibinfo {year} {2024})}\BibitemShut {NoStop}%
\bibitem [{\citenamefont {Ehrich}\ \emph {et~al.}(2020)\citenamefont {Ehrich}, \citenamefont {Esposito}, \citenamefont {Barra},\ and\ \citenamefont {Parrondo}}]{ehrich_micro-reversibility_2020}%
  \BibitemOpen
  \bibfield  {author} {\bibinfo {author} {\bibfnamefont {J.}~\bibnamefont {Ehrich}}, \bibinfo {author} {\bibfnamefont {M.}~\bibnamefont {Esposito}}, \bibinfo {author} {\bibfnamefont {F.}~\bibnamefont {Barra}},\ and\ \bibinfo {author} {\bibfnamefont {J.~M.}\ \bibnamefont {Parrondo}},\ }\href {https://doi.org/10.1016/j.physa.2019.122108} {\bibfield  {journal} {\bibinfo  {journal} {Phys. A: Stat. Mech. Appl.}\ }\textbf {\bibinfo {volume} {552}},\ \bibinfo {pages} {122108} (\bibinfo {year} {2020})}\BibitemShut {NoStop}%
\bibitem [{\citenamefont {De~Chiara}\ \emph {et~al.}(2018)\citenamefont {De~Chiara}, \citenamefont {Landi}, \citenamefont {Hewgill}, \citenamefont {Reid}, \citenamefont {Ferraro}, \citenamefont {Roncaglia},\ and\ \citenamefont {Antezza}}]{de2018reconciliation}%
  \BibitemOpen
  \bibfield  {author} {\bibinfo {author} {\bibfnamefont {G.}~\bibnamefont {De~Chiara}}, \bibinfo {author} {\bibfnamefont {G.}~\bibnamefont {Landi}}, \bibinfo {author} {\bibfnamefont {A.}~\bibnamefont {Hewgill}}, \bibinfo {author} {\bibfnamefont {B.}~\bibnamefont {Reid}}, \bibinfo {author} {\bibfnamefont {A.}~\bibnamefont {Ferraro}}, \bibinfo {author} {\bibfnamefont {A.~J.}\ \bibnamefont {Roncaglia}},\ and\ \bibinfo {author} {\bibfnamefont {M.}~\bibnamefont {Antezza}},\ }\href@noop {} {\bibfield  {journal} {\bibinfo  {journal} {NJP}\ }\textbf {\bibinfo {volume} {20}},\ \bibinfo {pages} {113024} (\bibinfo {year} {2018})}\BibitemShut {NoStop}%
\bibitem [{\citenamefont {Rodrigues}\ \emph {et~al.}(2019)\citenamefont {Rodrigues}, \citenamefont {De~Chiara}, \citenamefont {Paternostro},\ and\ \citenamefont {Landi}}]{rodrigues2019thermodynamics}%
  \BibitemOpen
  \bibfield  {author} {\bibinfo {author} {\bibfnamefont {F.~L.}\ \bibnamefont {Rodrigues}}, \bibinfo {author} {\bibfnamefont {G.}~\bibnamefont {De~Chiara}}, \bibinfo {author} {\bibfnamefont {M.}~\bibnamefont {Paternostro}},\ and\ \bibinfo {author} {\bibfnamefont {G.~T.}\ \bibnamefont {Landi}},\ }\href@noop {} {\bibfield  {journal} {\bibinfo  {journal} {PRL}\ }\textbf {\bibinfo {volume} {123}},\ \bibinfo {pages} {140601} (\bibinfo {year} {2019})}\BibitemShut {NoStop}%
\bibitem [{Note1()}]{Note1}%
  \BibitemOpen
  \bibinfo {note} {Given a characteristic length scale $R$ of the system-gas interaction, the low-density limit demands that $NR^3/V \ll 1$.}\BibitemShut {Stop}%
\bibitem [{\citenamefont {Taylor}(2012)}]{taylor2012scattering}%
  \BibitemOpen
  \bibfield  {author} {\bibinfo {author} {\bibfnamefont {J.}~\bibnamefont {Taylor}},\ }\href {https://books.google.de/books?id=OIaXvuwZMLQC} {\emph {\bibinfo {title} {Scattering Theory: The Quantum Theory of Nonrelativistic Collisions}}},\ Dover Books on Engineering\ (\bibinfo  {publisher} {Dover Publications},\ \bibinfo {year} {2012})\BibitemShut {NoStop}%
\bibitem [{\citenamefont {Blum}\ and\ \citenamefont {Gelbwaser-Klimovsky}(2024)}]{blum2024thermalization}%
  \BibitemOpen
  \bibfield  {author} {\bibinfo {author} {\bibfnamefont {S.}~\bibnamefont {Blum}}\ and\ \bibinfo {author} {\bibfnamefont {D.}~\bibnamefont {Gelbwaser-Klimovsky}},\ }\href@noop {} {\bibfield  {journal} {\bibinfo  {journal} {arXiv preprint arXiv:2409.15991}\ } (\bibinfo {year} {2024})}\BibitemShut {NoStop}%
\bibitem [{\citenamefont {Spohn}\ and\ \citenamefont {Lebowitz}(1978)}]{spohn2007irreversible}%
  \BibitemOpen
  \bibfield  {author} {\bibinfo {author} {\bibfnamefont {H.}~\bibnamefont {Spohn}}\ and\ \bibinfo {author} {\bibfnamefont {J.~L.}\ \bibnamefont {Lebowitz}},\ }\href {https://doi.org/https://doi.org/10.1002/9780470142578.ch2} {\bibfield  {journal} {\bibinfo  {journal} {Adv. Chem. Phys}\ }\textbf {\bibinfo {volume} {38}},\ \bibinfo {pages} {109} (\bibinfo {year} {1978})}\BibitemShut {NoStop}%
\bibitem [{\citenamefont {Alicki}(1979)}]{Alicki_1979}%
  \BibitemOpen
  \bibfield  {author} {\bibinfo {author} {\bibfnamefont {R.}~\bibnamefont {Alicki}},\ }\href {https://doi.org/10.1088/0305-4470/12/5/007} {\bibfield  {journal} {\bibinfo  {journal} {J. Phys. A Math. Theor.}\ }\textbf {\bibinfo {volume} {12}},\ \bibinfo {pages} {L103} (\bibinfo {year} {1979})}\BibitemShut {NoStop}%
\bibitem [{\citenamefont {Sakurai}\ and\ \citenamefont {Napolitano}(2020)}]{sakurai2020modern}%
  \BibitemOpen
  \bibfield  {author} {\bibinfo {author} {\bibfnamefont {J.~J.}\ \bibnamefont {Sakurai}}\ and\ \bibinfo {author} {\bibfnamefont {J.}~\bibnamefont {Napolitano}},\ }\href@noop {} {\emph {\bibinfo {title} {Modern quantum mechanics}}}\ (\bibinfo  {publisher} {Cambridge University Press},\ \bibinfo {year} {2020})\BibitemShut {NoStop}%
\bibitem [{\citenamefont {Shu}\ \emph {et~al.}(2019)\citenamefont {Shu}, \citenamefont {Cai}, \citenamefont {Seah}, \citenamefont {Nimmrichter},\ and\ \citenamefont {Scarani}}]{shu2019almost}%
  \BibitemOpen
  \bibfield  {author} {\bibinfo {author} {\bibfnamefont {A.}~\bibnamefont {Shu}}, \bibinfo {author} {\bibfnamefont {Y.}~\bibnamefont {Cai}}, \bibinfo {author} {\bibfnamefont {S.}~\bibnamefont {Seah}}, \bibinfo {author} {\bibfnamefont {S.}~\bibnamefont {Nimmrichter}},\ and\ \bibinfo {author} {\bibfnamefont {V.}~\bibnamefont {Scarani}},\ }\href@noop {} {\bibfield  {journal} {\bibinfo  {journal} {Phys. Rev. A}\ }\textbf {\bibinfo {volume} {100}},\ \bibinfo {pages} {042107} (\bibinfo {year} {2019})}\BibitemShut {NoStop}%
\bibitem [{\citenamefont {Prositto}\ \emph {et~al.}(2025{\natexlab{b}})\citenamefont {Prositto}, \citenamefont {Forbes},\ and\ \citenamefont {Segal}}]{prositto2025equilibrium}%
  \BibitemOpen
  \bibfield  {author} {\bibinfo {author} {\bibfnamefont {A.}~\bibnamefont {Prositto}}, \bibinfo {author} {\bibfnamefont {M.}~\bibnamefont {Forbes}},\ and\ \bibinfo {author} {\bibfnamefont {D.}~\bibnamefont {Segal}},\ }\href@noop {} {\bibfield  {journal} {\bibinfo  {journal} {arXiv preprint arXiv:2501.05392}\ } (\bibinfo {year} {2025}{\natexlab{b}})}\BibitemShut {NoStop}%
\bibitem [{\citenamefont {Allahverdyan}\ \emph {et~al.}(2004)\citenamefont {Allahverdyan}, \citenamefont {Balian},\ and\ \citenamefont {Nieuwenhuizen}}]{Allahverdyan_2004}%
  \BibitemOpen
  \bibfield  {author} {\bibinfo {author} {\bibfnamefont {A.~E.}\ \bibnamefont {Allahverdyan}}, \bibinfo {author} {\bibfnamefont {R.}~\bibnamefont {Balian}},\ and\ \bibinfo {author} {\bibfnamefont {T.~M.}\ \bibnamefont {Nieuwenhuizen}},\ }\href {https://doi.org/10.1209/epl/i2004-10101-2} {\bibfield  {journal} {\bibinfo  {journal} {Europhysics Letters}\ }\textbf {\bibinfo {volume} {67}},\ \bibinfo {pages} {565} (\bibinfo {year} {2004})}\BibitemShut {NoStop}%
\bibitem [{\citenamefont {Reed}\ and\ \citenamefont {Simon}(1979)}]{reed1979iii}%
  \BibitemOpen
  \bibfield  {author} {\bibinfo {author} {\bibfnamefont {M.}~\bibnamefont {Reed}}\ and\ \bibinfo {author} {\bibfnamefont {B.}~\bibnamefont {Simon}},\ }\href@noop {} {\emph {\bibinfo {title} {III: scattering theory}}},\ Vol.~\bibinfo {volume} {3}\ (\bibinfo  {publisher} {Elsevier},\ \bibinfo {year} {1979})\BibitemShut {NoStop}%
\bibitem [{\citenamefont {Dollard}(1964)}]{dollard1964asymptotic}%
  \BibitemOpen
  \bibfield  {author} {\bibinfo {author} {\bibfnamefont {J.~D.}\ \bibnamefont {Dollard}},\ }\href@noop {} {\bibfield  {journal} {\bibinfo  {journal} {Journal of Mathematical Physics}\ }\textbf {\bibinfo {volume} {5}},\ \bibinfo {pages} {729} (\bibinfo {year} {1964})}\BibitemShut {NoStop}%
\bibitem [{\citenamefont {Morchio}\ and\ \citenamefont {Strocchi}(2016)}]{morchio2016dynamics}%
  \BibitemOpen
  \bibfield  {author} {\bibinfo {author} {\bibfnamefont {G.}~\bibnamefont {Morchio}}\ and\ \bibinfo {author} {\bibfnamefont {F.}~\bibnamefont {Strocchi}},\ }\href@noop {} {\bibfield  {journal} {\bibinfo  {journal} {Reviews in Mathematical Physics}\ }\textbf {\bibinfo {volume} {28}},\ \bibinfo {pages} {1650001} (\bibinfo {year} {2016})}\BibitemShut {NoStop}%
\end{thebibliography}
\end{document}